\documentclass[12pt,preprint]{aastex}

\def\ll{\left \langle}
\def\rr{\right \rangle}
\def\al{\left |}
\def\ar{\right |}
\def\k{{mn}}
\def\kk{{m,n}}
\def\LH{LH}
\newcommand{\ve}[1]{\mathbf{#1}}

\shorttitle{Mass distribution of M87}
\shortauthors{Wu \& Tremaine}

\begin{document}

\title{Deriving the mass distribution of M87 from globular clusters}
\author{Xiaoan Wu and Scott Tremaine}
\affil{Princeton University Observatory, Peyton Hall}
\affil{Princeton, NJ 08544-1001, USA}
\email{xawn@astro.princeton.edu, tremaine@astro.princeton.edu}

\begin{abstract}

   We describe a maximum-likelihood method for determining the mass distribution in
spherical stellar systems from the radial velocities of a population of discrete test
particles. The method assumes a parametric form for the mass distribution
and a non-parametric two-integral distribution function. We apply the method to a sample of 161 globular clusters 
in M87. We find 
that the mass within 32 kpc is $(2.4\pm0.6)\times 10^{12}\,$M${_\odot}$, and 
 the exponent of the density profile $\rho\propto r^{-\alpha}$ in the range $10-100$ kpc is $\alpha=1.6\pm0.4$. 
The energy distribution suggests a few kinematically 
distinct groups of globular clusters. The
anisotropy of the globular-cluster velocity distribution cannot be determined 
reliably with the present data. 
Models fitted to an NFW potential yield similar mass estimates but cannot constrain
the concentration radius $r_c$ in the range $10-500$ kpc.

\end{abstract}


\section{Introduction}
\label{sec:intro}

The mass distribution in elliptical galaxies is difficult to measure, in part
because ellipticals do not contain gas or star disks like those found in spiral
galaxies. The mass distribution is particularly uncertain beyond the effective
radius, where the surface brightness fades rapidly. Fortunately, 
globular clusters (GCs) are found at large distances in many elliptical galaxies. 
For example, more than $8\times10^3$ GCs are believed to be present in M87 outside 3 times the effective
radius $R_e\sim 7$ kpc \citep{Mcl99, Zei93}. GCs in nearby galaxies provide
relatively bright pointlike sources for which radial velocities can be determined.
By analyzing the statistics of the positions and radial velocities of GCs, we may constrain
the mass profiles of their host galaxies and 
explore the dark matter distribution in the outer parts of elliptical galaxies.
Planetary nebulae also provide excellent kinematic probes for similar reasons \citep[e.g.][]{Dou02}.

The database of extragalactic GCs with radial velocities has been steadily growing in recent years. 
The most intensively
observed systems are NGC 1399 \citep{Dir04} and M87. In this paper we examine the latter
system. \citet{Har86} and \citet{Mcl99} observed the density profile of GCs in M87 out to 110 kpc. \citet{Mou90} compiled 
a list of 43 GCs with radial velocities, 
from which \citet{MT93} derived a mass of ${5-10\times10^{12}}$ M$_{\odot}$ within 50 kpc. 
They also estimated that about 200 and 1000 GCs would be required to derive accurately the slope
of the mass profile of the galaxy and the velocity anisotropy of the GCs, respectively. 
\citet{Coh97} and \citet{Coh00} obtained radial velocities for 221 GCs.
This dataset was used to study the spin \citep{Coh97, Coh98, Kis98} and
mass profile of M87 \citep{Rom01}.

Although there is significant rotation in the M87 GC system, the fractional 
kinetic energy in rotation
$(\Omega R/\sigma)^2\sim 20\%$ is relatively small, so it should be safe to ignore rotation
in our analysis and adopt the assumption that the system is spherical \citep{Cot01}.
Thus the data pairs $[R, v_z]$ contain all available dynamical information, where $R$ is
the projected distance from the GC to the center of the galaxy, and $v_z$ is its velocity in the line of sight.
To convert angular distance to physical distance, we adopt a distance to M87 of 
16.3 Mpc (i.e. 79 pc arcsec$^{-1}$) 
following \citet{Coh00}. 

   Maximum likelihood analysis provides a powerful tool to analyze discrete data. 
It was used to derive the mass distribution of spherical galaxies first by
\citet{Mer93} and \citet{MS93}. 
We describe our method for deriving spherical mass distributions from discrete data
in Section \ref{sec:tec}.
We use simulated data to test our method in
Section \ref{sec:sim}. In Section \ref{sec:M87} we constrain the
mass profile of M87 and the distribution function (hereafter DF)
of its GCs. The mass distribution of M87 has been determined from GC kinematics by several
authors, including \citet{MT93}, \citet{Coh97} and \citet{Rom01}. 
The study most similar to ours is that of  \citet{Rom01}. We compare
our method and results to theirs in Section \ref{sec:dis}. Section \ref{sec:con} 
contains conclusions.

\section{Methods}

\label{sec:tec}

We wish to constrain the potential $\Phi(r)$ of a spherical system from a set of discrete data pairs $[R_i, v_{zi}]$.
In a spherical system, the DF depends on at most two integrals of motion, 
energy $E$ and scalar angular momentum $L$, and thus 
takes the form $f(E, L)$. We normalize so that
\begin{eqnarray}
\int f(E,L)d^3{\ve{x}} d^3\ve{v}=1.
\end{eqnarray}
We introduce a Cartesian
coordinate system $(x, y, z)$ with origin at the galactic center, and $z$-axis along the line of sight. 
In this system, the projected radius is
\begin{eqnarray}
R=\sqrt{x^2+y^2}. 
\end{eqnarray}
The probability that a given GC is found in the interval $dRdv_z$ is 
$2\pi Rg(R, v_z)dRdv_z$, where
\begin{eqnarray}
g(R, v_z)=\int f(E, L) dz dv_x dv_y. \label{eq:gRvz}
\end{eqnarray}
Because of spherical symmetry, we may assume without losing any information that
all GCs lie in the $x-z$ plane and $x>0$.
Thus we have
\begin{eqnarray}
E&=&\Phi(r)+\frac{1}{2}(v_x^2+v_y^2+v_z^2), \nonumber\\ 
L&=&\sqrt{(v_zR-v_xz)^2+v_y^2(R^2+z^2)}, \label{eq:EL3D}
\end{eqnarray}

Given a potential $\Phi(r)$ and a DF $f(E, L)$, the likelihood of observing the
data pairs $[R_i, v_{zi}]$ is
\begin{eqnarray}
\LH(\Phi, f)& \equiv &\LH([R_i, v_{zi}]|\Phi, f) \nonumber \\
   &=&\prod _{i} g(R_i, v_{zi}|\Phi, f) \cdot 2\pi R_i \nonumber\\
   &\propto& \prod _{i} g(R_i, v_{zi}|\Phi, f).
\end{eqnarray}
The maximum likelihood method assumes that {\it{the best estimate of the
potential and DF is the one that maximizes 
$\LH(\Phi, f)$}}. More generally, Bayes' theorem states that 
if we know some prior information $H$ 
which gives a prior probability $P(\Phi, f|H)$, we should maximize the posterior
probability $P(\Phi, f|H)\LH(\Phi, f)$ rather than the likelihood.

This formalism can easily be generalized to account for selection effects and observational
errors in radial velocities.
For example, in most surveys
we observe GCs only in a limited range of radii
$(R_{s0}<R<R_{s1})$. To account for this limitation, we simply modify the normalization to require that
\begin{eqnarray}
\int ^{R_{s1}} _{R_{s0}} g(R, v_z) 2\pi RdR\int dv_z =1. \label{eq:gRvzN}
\end{eqnarray}
To account for the errors in radial velocities, 
we just need to calculate $\LH$ with $g(R, v_z)$ replaced by
\begin{eqnarray}
g'(R, v_z)=\frac{1}{\sqrt{2\pi}\sigma_v}\int g(R, v_z) e^{\displaystyle{-(v-v_z)^2/2\sigma_v^2}}dv. \label{eq:error}
\end{eqnarray}
Here we have assumed that the error $\sigma_v$ is the same for all GCs, although the method is easily generalized to individual measurement errors.
Note that our method also probes, though with less power, the potential beyond the maximum survey radius $R_{s1}$,
both because we observe the projected distribution, which includes GCs with radial distances to the galactic center
larger than $R_{s1}$, and because some GC orbits have apocenters outside $R_{s1}$.

In principle, all physically possible potentials and DFs should be explored to
maximize $\LH(\Phi,f)$. The most general way to do so is to assume non-parametric forms for both potentials
and DFs, but this is cumbersome and computationally expensive. Instead we optimize within
a family of parametrized potentials: we assume an analytical form $\Phi(r, \ve{X})$
and infer the parameters $\ve{X}$ (which can be viewed as a row vector). For example, we may assume
a power-law density-potential pair 
\begin{eqnarray}
\rho(r)&=&\rho_0\left (\frac {r}{r_0}\right )^{-\alpha}, \nonumber\\
\Phi(r)&=&\frac{4\pi G\rho_0 r_0^2}{(2-\alpha)(3-\alpha)}\left (\frac {r}{r_0} \right )^{2-\alpha} \;\;\;(\alpha<3),  \label{eq:powerlaw}
\end{eqnarray}
where $r_0$ is an arbitrary radius,
and then $\ve{X}={\{\rho_0, \alpha\}}$. We also investigate
the \cite{Nav97} density-potential pair,
\begin{eqnarray}
\rho(r)&=&\frac{\rho_0}{\displaystyle{\frac{r}{r_c}}\left (1+\displaystyle{\frac{r}{r_c}} \right)^2}, \nonumber\\
\Phi(r)&=&-4\pi G\rho_0r_c^2\frac{\ln \left (1+\displaystyle{\frac{r}{r_c}} \right)}{\displaystyle{\frac{r}{r_c}}}. \label{eq:NFW}
\end{eqnarray}
Here $r_c$ is the concentration radius and $\ve{X}={\{\rho_0, r_c\}}$.

If the DF $f(E, L)$ were known, the problem of optimizing the parameters $\ve{X}$ would be easy to solve. 
We would calculate $\LH$ as a function of $\ve{X}$ and figure out where $\LH$ peaks.
However, in practice the DF is an unknown function,
and we have to derive the DF and $\ve{X}$ simultaneously.
This can be done by parametric or non-parametric methods. 

In the parametric method \citep{MS93}, we choose a limited number of 
basis functions $f_k(E, L)$ (i.e. a complete set of basis functions truncated to some order $N$), 
which are smooth in phase space. Thus, the DF is approximated as 
\begin{eqnarray}
f(E, L)=\sum _{k} w_kf_k(E, L), & & k=1, ..., N. \label{fkE}
\end{eqnarray}
This method ensures that the derived DF is smooth; however, it 
is difficult to ensure that $f(E, L)\geq 0$ for all $(E, L)$ since neither $w_k$ nor $f_k$ is necessarily
non-negative. Furthermore, if $N$ is small
we cannot ensure that equation (\ref{fkE}) gives a good approximation to the actual DF.
Of course, we can always increase $N$ to improve the approximation,
but then we cannot ensure that the best-fit DF will be smooth.

In the non-parametric method \citep{Mer93}, we divide the $E-L$ space into 
$N_E\times N_L$ bins, which are denoted by the double index $\k$, $m=1, ..., N_E$,
$n=1, ..., N_L$. Notice that the DF is isotropic if $N_L=1$ and otherwise may be anisotropic. 
Then we construct a set of top-hat functions, 
\begin{eqnarray}
h_\k(E,L)=\left \{{
        \begin{array}{rl}
        {V_\k^{-1}} & \hbox{if $V_\k\neq 0$ and $(E,L)\in$ bin $\k$} \\
        0 & \hbox{otherwise}, \label{fkEstep} \\
        \end{array}} \right.
\end{eqnarray}
where $V_\k\equiv\int _{\hbox{bin $\k$}}d^3\ve{x}d^3\ve{v}$ is the phase-space volume within the survey limits that is associated with bin $\k$.
Thus, we have
\begin{eqnarray}
\int h_\k(E, L)d^3\ve{x}d^3\ve{v} &=& 1, \\
f(E, L) &=&\sum _\kk w_\k h_\k(E, L),\label{fEL}\\
f_\k&=&\left \{{
             \begin{array}{rl}
             {{w_\k}/{V_\k}}  \hbox{$\;\;\;\;$if $V_\k\neq 0$} \\
              0  \hbox{$\;\;\;\;$otherwise}, \\
             \end{array}} \right. \\
\sum _\kk w_\k&=&1, \\
\ve{W}&\equiv& \{w_\k\}. \label{eq:W}
\end{eqnarray}
Here $w_\k$ is simply the fraction of GCs in bin $\k$, $f_\k$ is the phase space density of
GCs in bin $\k$, and 
$f(E, L)$ and $h_\k(E, L)$ are normalized so that their integrals over phase space within
the survey limits are unity.

In our calculation, the bin $\k$ is defined by
\begin{eqnarray}
E_{m-1}&\leq E <& E_{m},\nonumber \\
\frac{n-1}{N_L}L_c(E_{m}) &\leq L <&\frac{n}{N_L}L_c(E_{m}), \label{eq:EL}
\end{eqnarray}
where $E_{m-1}$ and $E_{m}$ are the lower and upper energy limits of the bin, and $L_c(E)$ is the
maximum angular momentum at energy $E$, corresponding to a circular orbit.
These bins cover all of the allowed region in $E-L$ space and may also cover some unallowed regions.

Each $h_\k(E, L)$ gives a normalized distribution in $(R, v_z)$ space,
\begin{eqnarray}
g_\k(R, v_z)=\int h_\k(E, L) dv_x dv_y dz. \label{eq:gkRvz}
\end{eqnarray}
Notice that $g_\k(R, v_z)$ depends on $\Phi$ (thus $\ve{X}$) but is independent of the DF specified by $\ve{W}$.
An algorithm for evaluating $g_\k(R, v_z)$ efficiently is shown
in Appendix \ref{sec:Appendix}. 
To account for errors in radial velocities, the distribution should be convolved with
a one-dimensional Gaussian function as in equation (\ref{eq:error}). We use the convolved
distribution hereafter and drop the prime symbol for simplicity. 

Due to the linearity of equations (\ref{eq:gRvz}) and (\ref{fEL}), we have
\begin{eqnarray}
g(R, v_{z})&=&\sum _\kk w_\k g_\k(R, v_{z}), \nonumber\\
\LH(\ve{X}, \ve{W})&\propto &\prod _i g(R_i, v_{zi}) \nonumber\\
                  &=&\prod _i\sum _\kk w_\k g_\k(R_i, v_{zi}). \label{eq:LH}
\end{eqnarray}
Therefore, for a given set of observations $[R_i, v_{zi}]$, the likelihood is a function of $\ve{X}$ and $\ve{W}$ only. 
For a fixed potential specified by the parameters $\ve{X}$, we can 
calculate $g_\k(R_i, v_{zi})$, then maximize $\LH(\ve{X}, \ve{W})$ with respect to $\ve{W}$. We may
then vary $\ve{X}$ to search for a global maximum $\LH(\ve{X}, \ve{W})$. 

An advantage of this method is that we need only ensure that all $w_\k$ are non-negative to 
guarantee that $f(E, L)\geq 0$. 
However, this method does not generally give smooth DFs. 
The derived DFs look like the sum of a set of $\delta$-functions \citep
{Mer93}.
This is a drawback,
since to some extent the irregularities in the DFs are fitting the statistical fluctuations in the data
rather than the signal. In other words it is an ill-posed problem to infer $g(R_i, v_{zi})$ from a small set
of data points.
One way to approach this problem is to maximize a ``penalized likelihood'' instead,
which can be defined using entropy maximization \citep{Ric88} or
regularization \citep{Mer93, Rix97}.  

By adding an entropy term, we maximize the function
\begin{eqnarray}
T&\equiv&\ln \LH+\alpha S, \label{eq:alpha} \\
S&\equiv& -\int C(f) d^3\ve{x}d^3\ve{v}, \label{eq:entropy}
\end{eqnarray}
where $\alpha$ is a parameter chosen to control the degree of smoothness of the DF, and
$C(f)$ is a convex function of $f$ such as $f\ln f$.
Adding the entropy term tends to force the DF to be uniform in phase space (in the
limit of large $\alpha$, the data become irrelevant and the maximum value of $T$ occurs
for $f=\hbox{const}$). This tendency is particularly undesirable in our case because there is
a large phase volume at high energy, so that large $\alpha$ tends to place more
GCs in a high energy state.
We have done simulations to test this method and found that
the derived DF is either too irregular for small $\alpha$ or has
an excess high energy tail for large $\alpha$. 

To implement regularization, we maximize the function 
\begin{eqnarray}
Q(\ve{X}, \ve{W})\equiv\ln \LH-\lambda_E\Pi_E-\lambda_L\Pi_L, \label{eq:Q} 
\end{eqnarray}
where $\lambda_E$ and $\lambda_L$ are positive parameters that adjust the degree of smoothness of the DF
and 
$\Pi_E$ and $\Pi_L$ are dimensionless positive-definite functions of the DF. There are several
natural forms for $\Pi_E$, including 
\begin{eqnarray}
\Pi_E &= \ll \Bigl ({\frac{\displaystyle{\partial \ln f}}{\displaystyle{\partial (E/E_{c})}}\Bigr )^2}\rr  &\hbox{uniform form},\label{eq:uniformDF} \\
\Pi_E &= \ll \al\frac{\displaystyle{\partial ^2\ln f}} {\displaystyle{\partial (E/E_{c})^2}}\ar\rr &\hbox{exponential form}, \label{eq:expformDF}\\
\Pi_E &= \ll \al\frac{\displaystyle{\partial ^2\ln f}} {\displaystyle{\partial \ln (E/E_{c})^2}}\ar\rr &\hbox{power-law form}, \label{eq:power-lawformDF}
\end{eqnarray}
where 
$E_{c}$ is a normalizing constant. The names indicate what functions minimize $\Pi_E$; for example,
the uniform form is zero if $f=\hbox{const}$, the exponential form is zero if $f\propto \exp(-\beta E)$, where $\beta$
is a constant, etc.
The angle brackets denote an average over area in $E-L/L_c(E)$ space;
for example, if we use a fixed energy interval for all bins,
the exponential form is proportional to
\begin{eqnarray}
 \sum _\kk \al{\ln f_{(m+1)n}-2\ln f_{mn}+\ln f_{(m-1)n}} \ar ,
\end{eqnarray}
in case no $V_\k$ is zero.
The term $\Pi_L$ smoothing the DF along the $L$ direction can be written similarly.

Before we choose the form of $\Pi_E$ and $\Pi_L$, it is helpful to discuss the meaning of the regularization term.
Equation (\ref{eq:Q}) is equivalent to
\begin{eqnarray}
e^Q&=&e^{-\lambda_E\Pi_E-\lambda_L\Pi_L}\LH(\Phi,f).
\end{eqnarray}
Therefore, maximizing $Q$ is equivalent to maximizing the posterior 
probability if the prior probability for $f$ is 
\begin{eqnarray}
P(f|H_f)\equiv e^{-\lambda_E\Pi_E-\lambda_L\Pi_L}.
\end{eqnarray}
By choosing $\lambda_E$ and $\Pi_E$, we are assuming some prior information $H_f$, i.e., a specific degree of smoothness and 
shape of the DF. Therefore, the smoothing terms assess the physical plausibility of the 
DF \citep{Cox90}.
 
Generally, DFs of GCs are strongly non-uniform in the sense that they may change amplitude
by a few orders of magnitude over the observable range. For example, given 
a logarithmic potential and a GC density profile
\begin{eqnarray}
\Phi=2\sigma^2\ln r, & &
\nu(r) \propto r^{-3},
\end{eqnarray}
the corresponding isotropic DF is 
\begin{eqnarray}
f(E)\propto e^{\displaystyle{-3E/2\sigma ^2}}.
\end{eqnarray} 
Over the range $R_0<R<R_1$, the DF at zero velocity will vary by a factor of
$(R_1/R_0)^3$, which is about 3000, given that the ratio $R_1/R_0$ is 110 kpc/7 kpc $\sim$ 15 in the case of M87 
(see \S\ref{sec:M87}).

Given this large variation, neither entropy maximization as in equation (\ref{eq:entropy}) nor
the uniform form of the regularization term in equation (\ref{eq:uniformDF}) is appropriate, 
because they tend to favor a uniform DF over the entire phase space.
The exponential and power-law forms in equations (\ref{eq:expformDF}) and (\ref{eq:power-lawformDF}) 
give more freedom
and favor exponential or power-law forms for
the DF, which are more physically reasonable.
We adopt the exponential form throughout this paper.

For convenience, we define 
\begin{eqnarray}
Q_{max}(\ve{X})\equiv \max_\ve{W} Q(\ve{W}, \ve{X}), \label{eq:Qw}
\end{eqnarray}
which is $Q$ maximized with respect to the weights $\ve{W}$.
The function $Q(\ve{W}, \ve{X})$ usually has a number of local maxima in $\ve{W}$ at a fixed set of potential parameters
$\ve{X}$. We attempt
to find the global maximum in this landscape using simulated annealing and the downhill simplex
method, as discussed in 
\citet{Pre03}. The method does not guarantee that we have found the global maximum, so we
try several different initial values of $\ve{W}$ and choose the largest maximum found from any of 
these initial conditions.
The optimization process turns out to consume most of the computational resources required by
this method.

We may also assume some prior information $H_X$ for the potential parameters, which gives a prior 
probability $P(\ve{X}|H_X)$. For example, in the power-law potential (eq. \ref{eq:powerlaw}), 
it is reasonable to assume a uniform prior in $(\ln \rho_0, \alpha)$, thus we have
\begin{eqnarray}
P(\ve{X}|H_X)=\rho_0^{-1}.
\end{eqnarray}
Then the probability distribution of the potential parameters $\ve{X}$
is 
\begin{eqnarray}
P(\ve{X}) \propto P(\ve{X}|H_X)e^{Q_{max}(\ve{X})} .
\end{eqnarray}

   We may also incorporate the observed surface number density profile $\Sigma_{obs}(R_j)$ of the GCs and 
its error $\sigma_\Sigma(R_j)$
as an additional constraint in our calculation since this may be available for many
more GCs, over a larger survey area than have measured velocities.
We maximize the quantity
\begin{eqnarray}
Q' &\equiv&\ln \LH-\frac{\chi ^2}{2}-\lambda_E\Pi_E-\lambda_L\Pi_L, \label{eq:Qwp}\\
\chi ^2&=&\sum_j \frac {(\Sigma(R_j)-\Sigma_{obs}(R_j))^2}{\sigma_\Sigma(R_j)^2}, \label{eq:x2}\\
\Sigma(R) &\propto &\int g(R, v_z) dv_z \nonumber \\
          &= &\sum_\kk w_\k \int g_\k(R, v_z) dv_z,
\end{eqnarray}
where both $\Sigma(R)$ and $\Sigma_{obs}(R)$ should be normalized so that $2\pi\int RdR\Sigma(R)$ and
$2\pi\int RdR\Sigma_{obs}(R)$ are unity within the survey limit of GC number counting.

How should we choose $\lambda_E$ and $\lambda_L$? If these parameters are set too small, 
they do not smooth the DF at all. If
they are set too large, the DF is forced to a specific functional form that may not be demanded
by the data. Notice that in contrast to many problems such as least-squares fitting \citep{Pre03} 
the likelihood is only 
known to within a multiplicative constant, so the goodness of fit can only be assessed in
a relative sense. 
When maximizing $Q'$ 
(eq. \ref{eq:Qwp}), we 
choose smoothing parameters following \citet{Rix97}. 
Assume that the minimum $\chi ^2$ in equation (\ref{eq:x2})
is $\chi_0 ^2$ in case of no regularization. Then we 
increase the smoothing parameters so that  
the minimum $\chi ^2$ is $\chi_0 ^2+n_s^2$, where $n_s$ corresponds to $n_s-\sigma$ error. 
We call these ``appropriate'' smoothing parameters. A degeneracy arises
in choosing two smoothing parameters from one requirement. The two parameters $\lambda_E$
and $\lambda_L$
control smoothness in the direction of $E$ and $L$, respectively and are
not necessarily equal. Fortunately, we
find that all ``appropriate'' pairs of smoothing parameters give similar results for
derived potentials and energy distributions, although, not surprisingly, the derived anisotropy can
be very different depending on the value of $\lambda_L$. Of course, with more data, the values
of smoothing parameters will be subject to tighter constraints.
Therefore, we may recover the anisotropy if enough data is available.

\section{Simulations}

\label{sec:sim}

  We first test our algorithm using simulated data. We choose 
a DF 
\begin{eqnarray}
f(E, L)=  \left (e^{-(E-320)^2/100}+0.02 e^{-(E-380)^2/100} \right ) e^{\gamma L/L_c(E)} \label{eq:powerlawDF}
\end{eqnarray}
in a power-law potential (eq. \ref{eq:powerlaw}) with $r_0=19$ kpc and 
$(\rho_0, \alpha) =(1.9\times10^7\,\hbox{M}_{\sun}\,\hbox{kpc}^{-3}, 1.9)$. 
The energy $E$ is in units of $(100\,\hbox{km s}^{-1})^2$.
The anisotropy parameter $\gamma$ is a constant which we choose in the range
from $-5$ (extremely radial) to 5 (extremely tangential).
The DF is chosen so that the energy distribution is bimodal, to challenge the algorithm.
We generate 160 kinematic data points within the projected radius range 
7--32 kpc and a surface number
density profile from 7000 GCs within the projected radius range 7--110 kpc (these match
the properties of the sample of M87 GCs described in \S\ref{sec:M87data}).

\begin{figure}
\epsscale{0.47}
\plotone{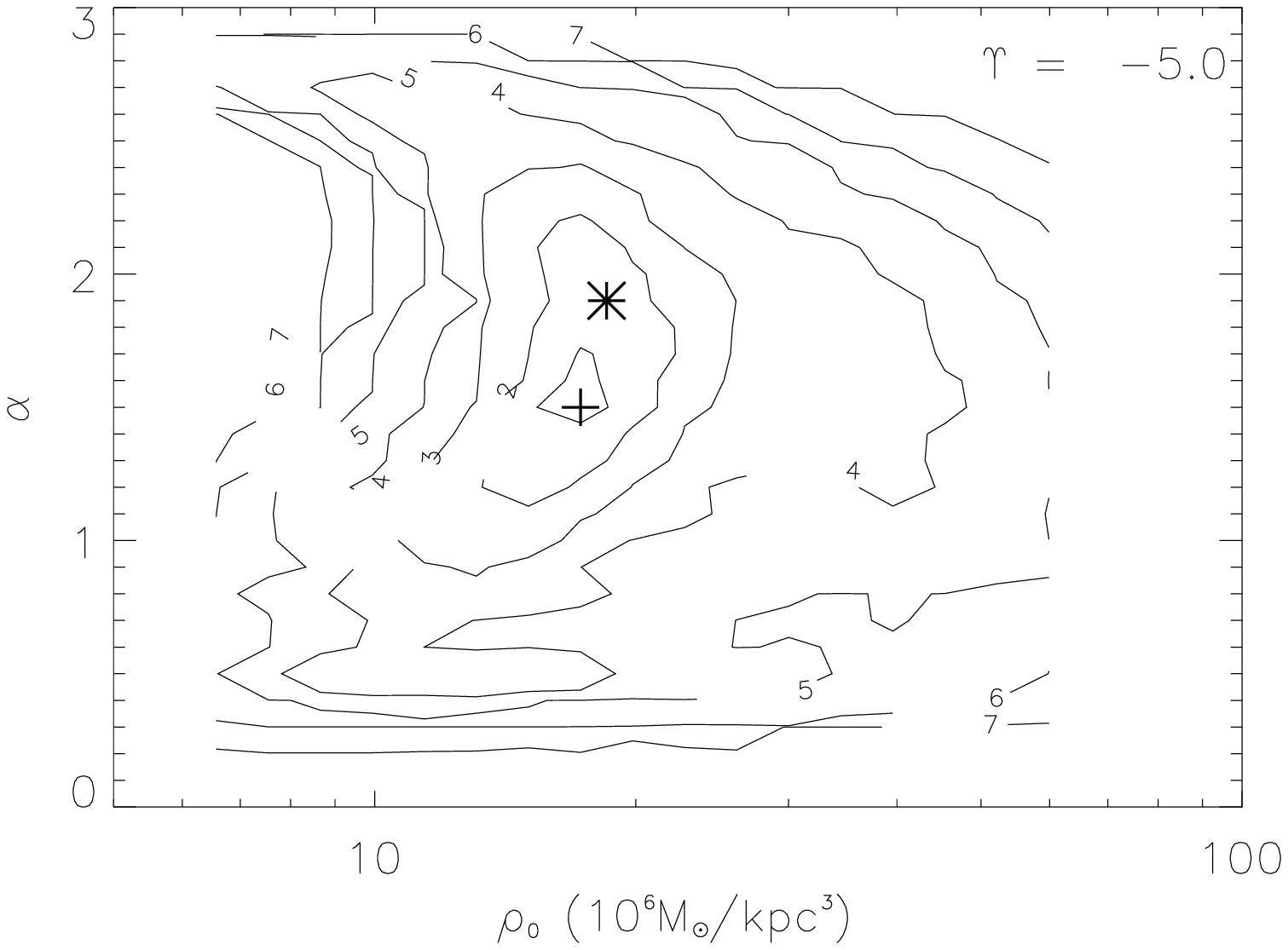}
\plotone{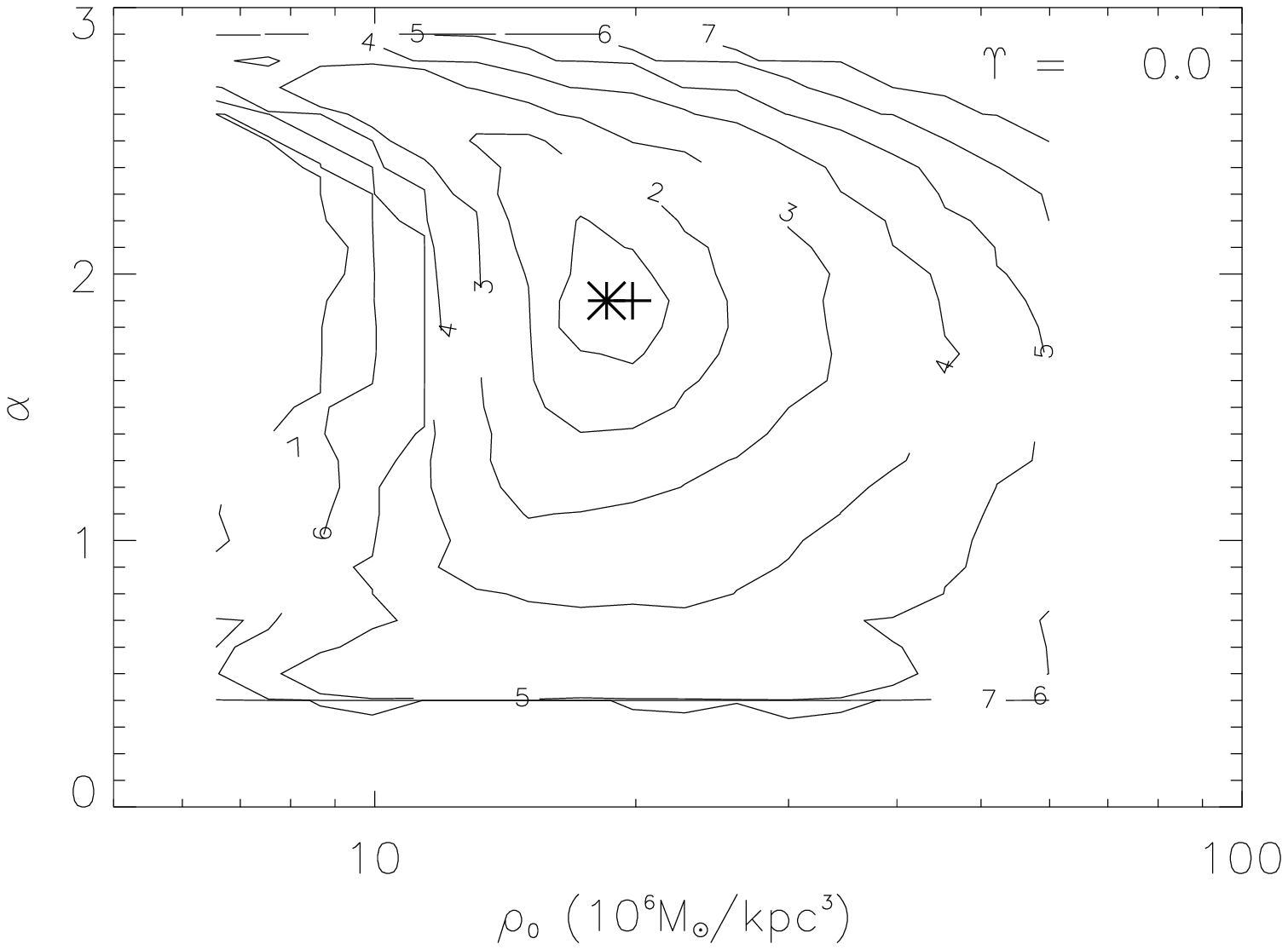}
\plotone{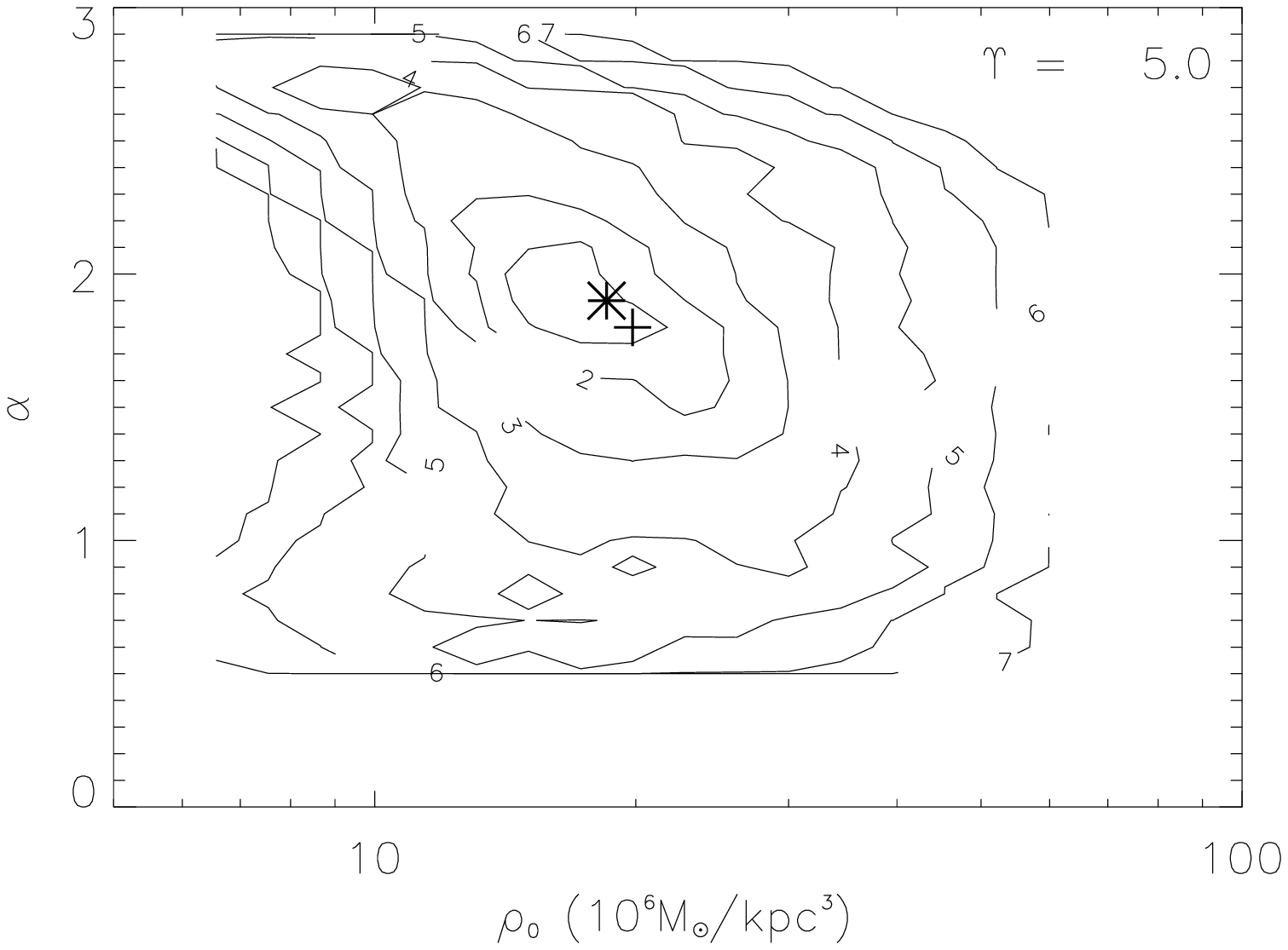}
\epsscale{1.0}
\figcaption{Tests of the recovery of the potential parameters  $\ve{X}=\{\rho_0, \alpha\}$ 
and the DFs, using 160 data points with radial velocities and
7000 data points to define the surface density profile in a
power-law potential (eq. \ref{eq:powerlaw}). The input parameters
are marked by the asterisk signs. 
The input DFs are described by equation (\ref{eq:powerlawDF}) 
with anisotropy parameter $\gamma=-5$ (radial), 0 (isotropic) and 5 (tangential).
Equation (\ref{eq:expformDF}) is used for regularization. 
The plus signs mark the best-fit model, which gives the maximum $Q'_{max}(\ve{X})$ (eq. \ref{eq:Qwp}).
The contours are $n-\sigma$ away from the peak. The tests show that we recover the potential parameters
pretty well even for extremely anisotropic DF.
\label{fig:fig1}}
\end{figure}

In this case, $\ve{X}=\{\rho_0, \alpha\}$.
Figure \ref{fig:fig1} shows the contour plot for $Q_{max}(\ve{X})$ 
for three simulations with $\gamma=-5$, 0 and 5. The appropriate
smoothing parameters are $(\lambda_E, \lambda_L)=(0.015, 0.015)$ so that $n_s=1$.
The plus signs are the best-fit model which gives the maximum $Q_{max}(\ve{X})$ while
the asterisk signs mark the input models.
The contours are $n-\sigma$ levels from the peak, i.e., $e^{-n^2/2}$ of the maximum likelihood.
We run 20 simulations with $\gamma$ between $-5$ and 5 and find that the standard errors
are 10\% for $\rho_0$ and 0.2 for 
$\alpha$. These tests show that we can recover the potential parameters from this dataset, even for
extremely anisotropic DFs.

\begin{figure}
\epsscale{0.47}
\plotone{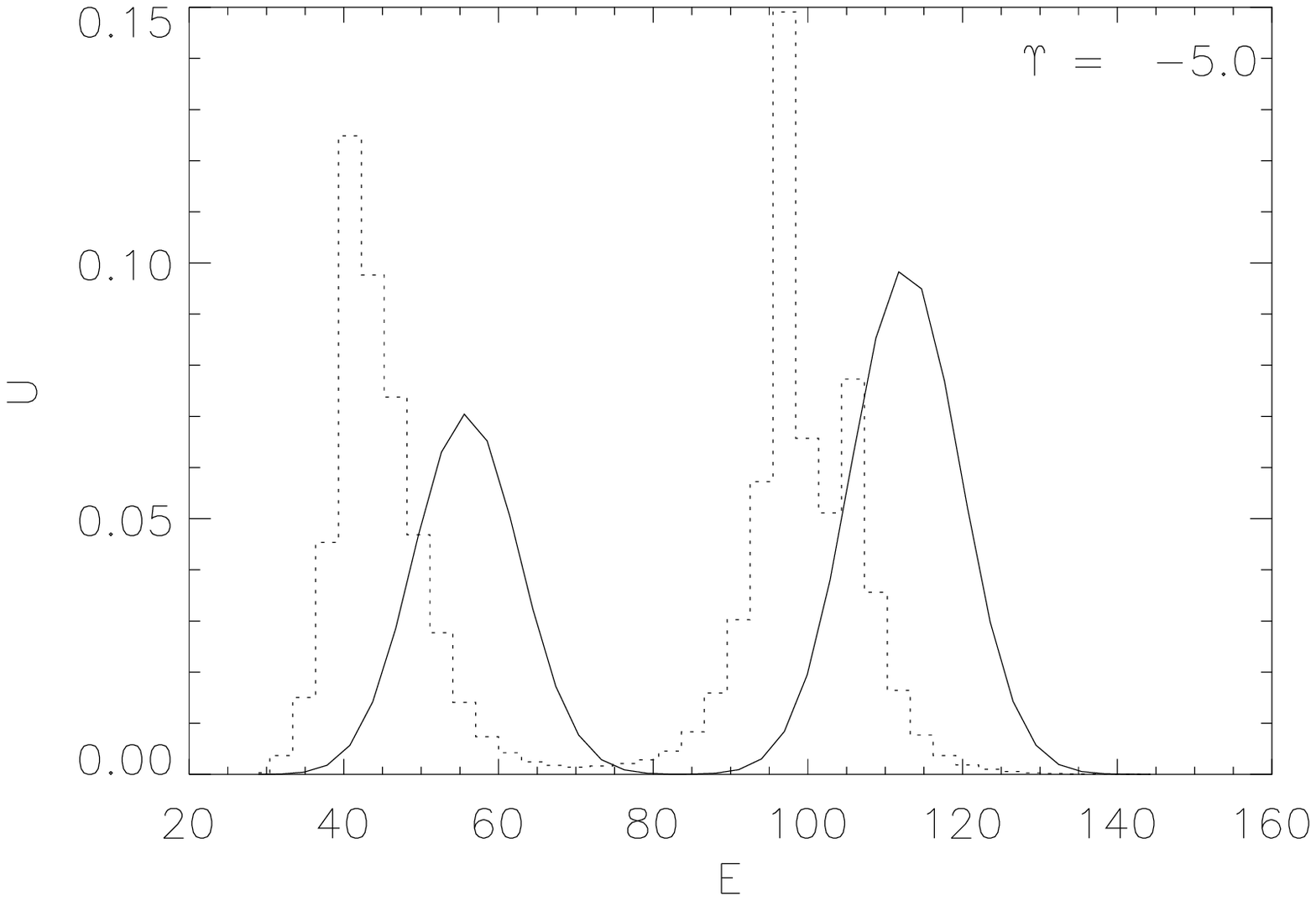}
\plotone{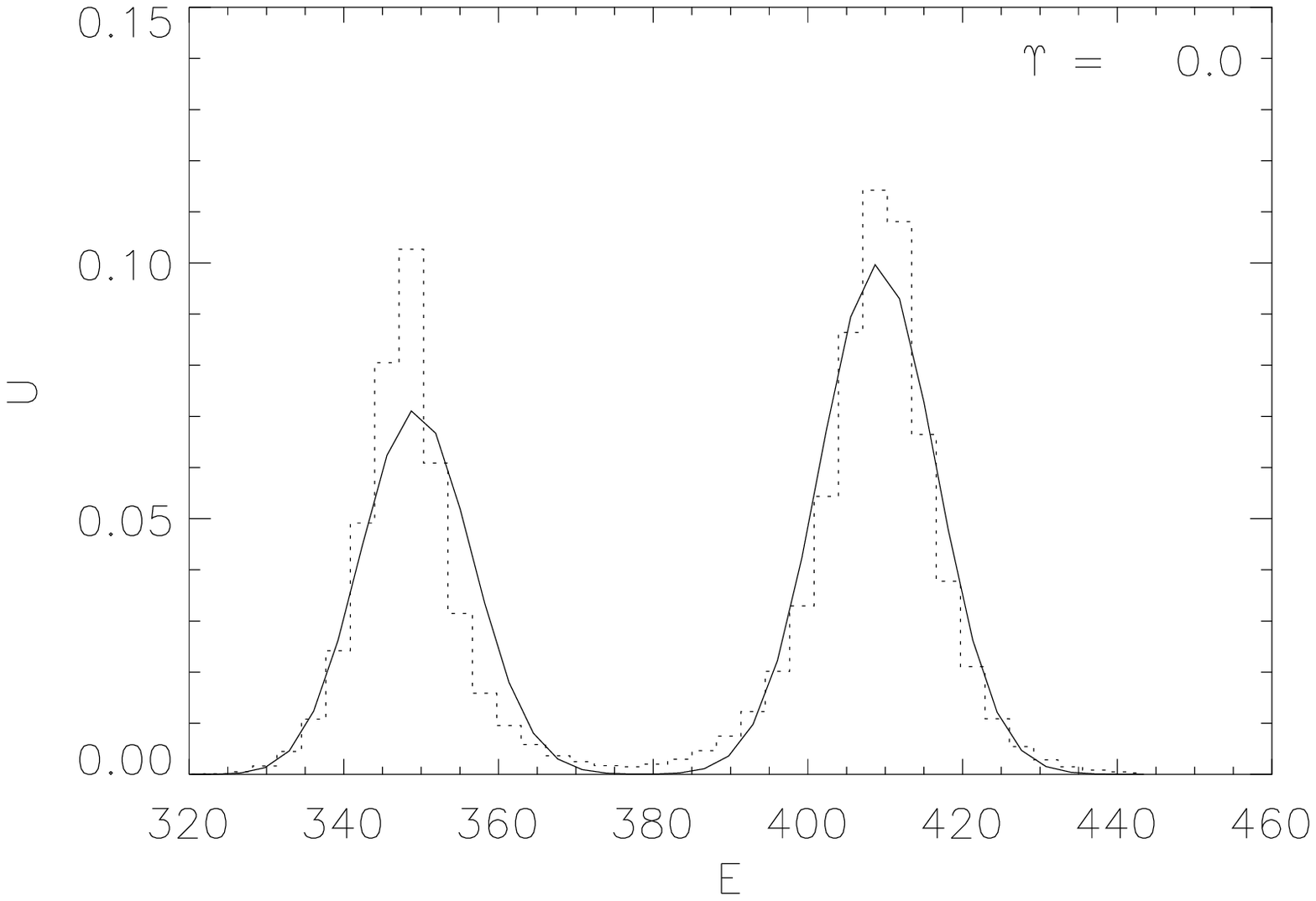}
\plotone{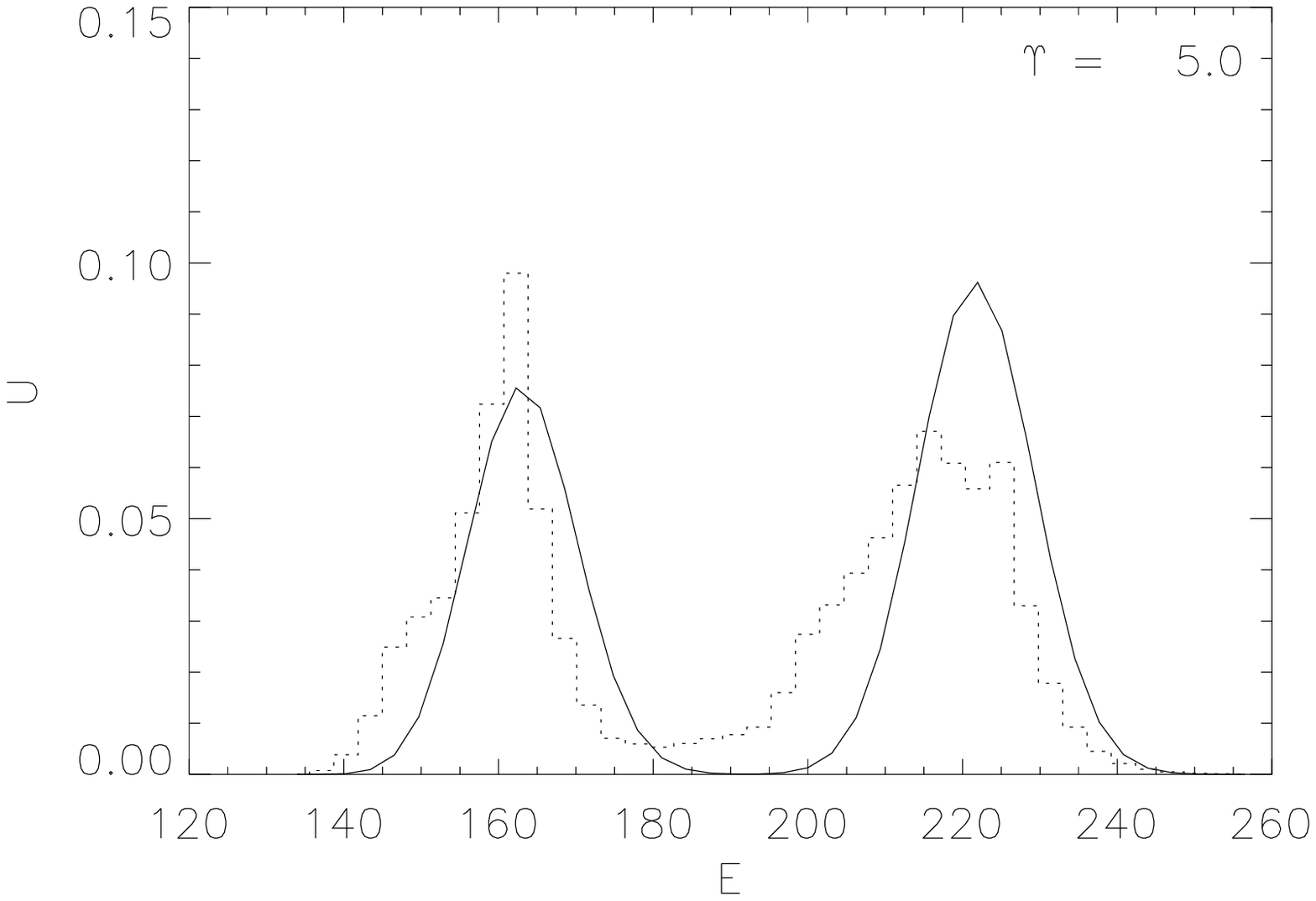}
\epsscale{1.0}
\figcaption{Derived energy distributions associated with the best-fit models in Figure 
\ref{fig:fig1}.
The derived energy distributions (dotted lines) successfully recover the bimodality of the input energy distributions 
(solid lines) for all cases. The shift between the input and derived
energy distributions for the radial DF ($\gamma=-5$) is related to the relatively large errors
in derived potential parameters, 
which shift the zero-energy point. 
 The results show that 
we can recover the energy distribution from a small ($\sim$160 velocities) data
set reasonably well. 
\label{fig:fig2}}
\end{figure}

Figure \ref{fig:fig2} shows the input and derived energy distributions of the GCs, described by
\begin{eqnarray}
\ve{U}\equiv \left \{\sum_n w_\k\right \},
\end{eqnarray}
i.e., the sum of the weights over all angular-momentum bins at a given energy.
The derived energy distributions (dotted lines) fit the input distributions (solid lines) 
reasonably well for the nearly radial ($\gamma=5$) and isotropic ($\gamma=0$) DFs.
However, the derived DF is shifted to low energies  for the nearly radial DF ($\gamma=-5$),
probably because the relatively large errors (see Figure \ref{fig:fig1}) 
in the derived potential parameters lead to
a shift of zero-energy point.
Nevertheless, the bimodality of the DF is successfully recovered.

\begin{figure}
\epsscale{0.47}
\plotone{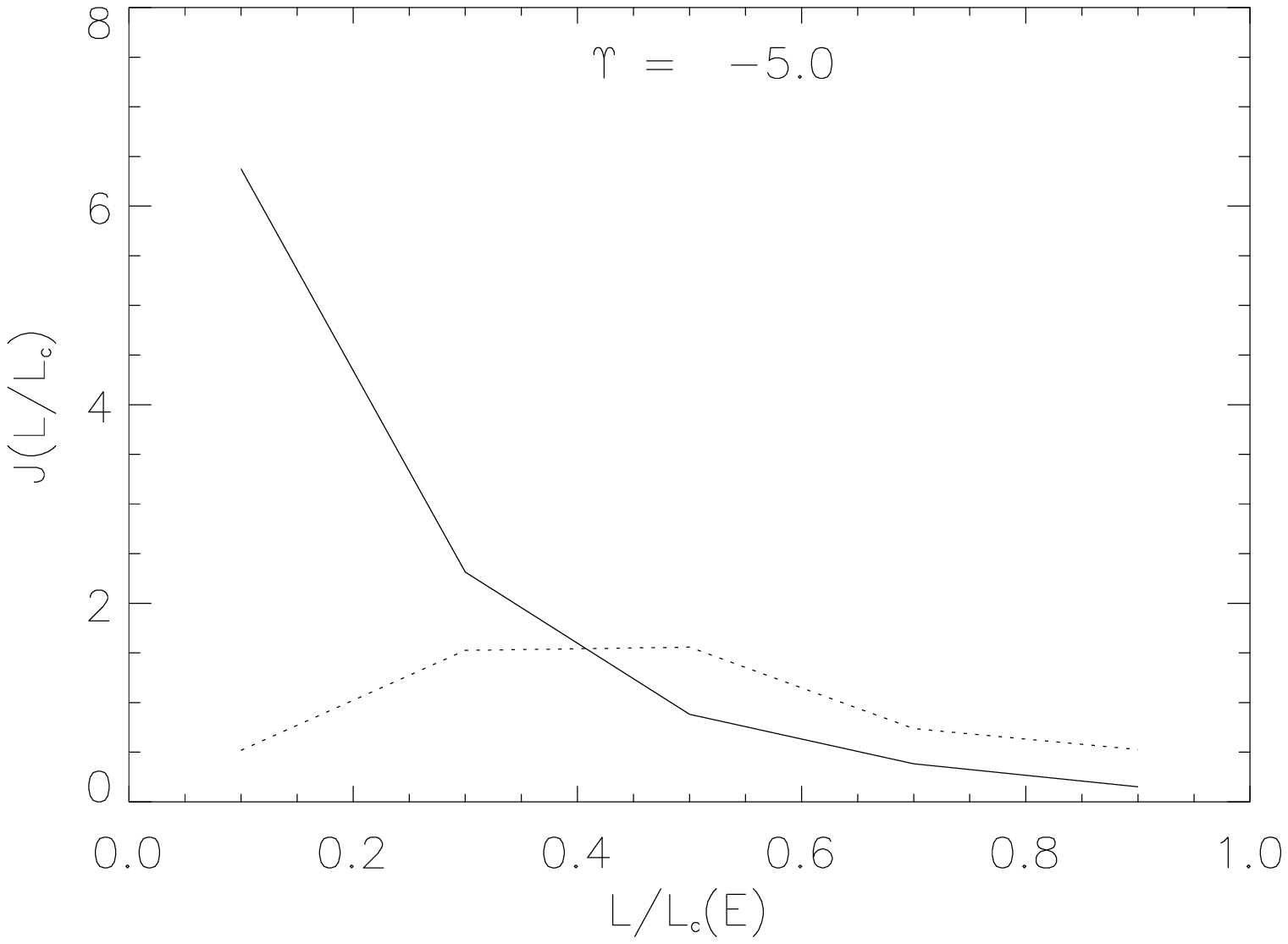}
\plotone{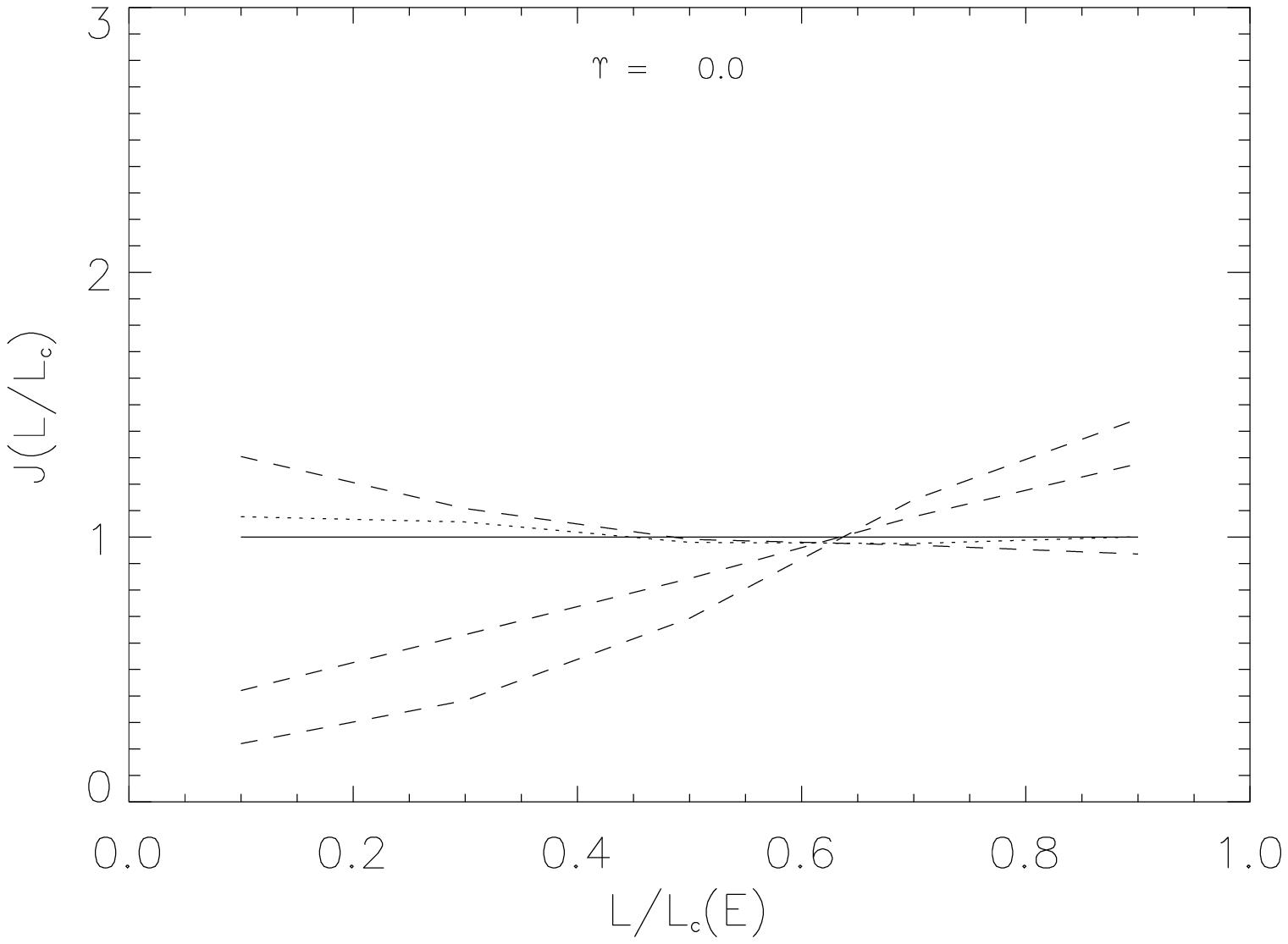}
\plotone{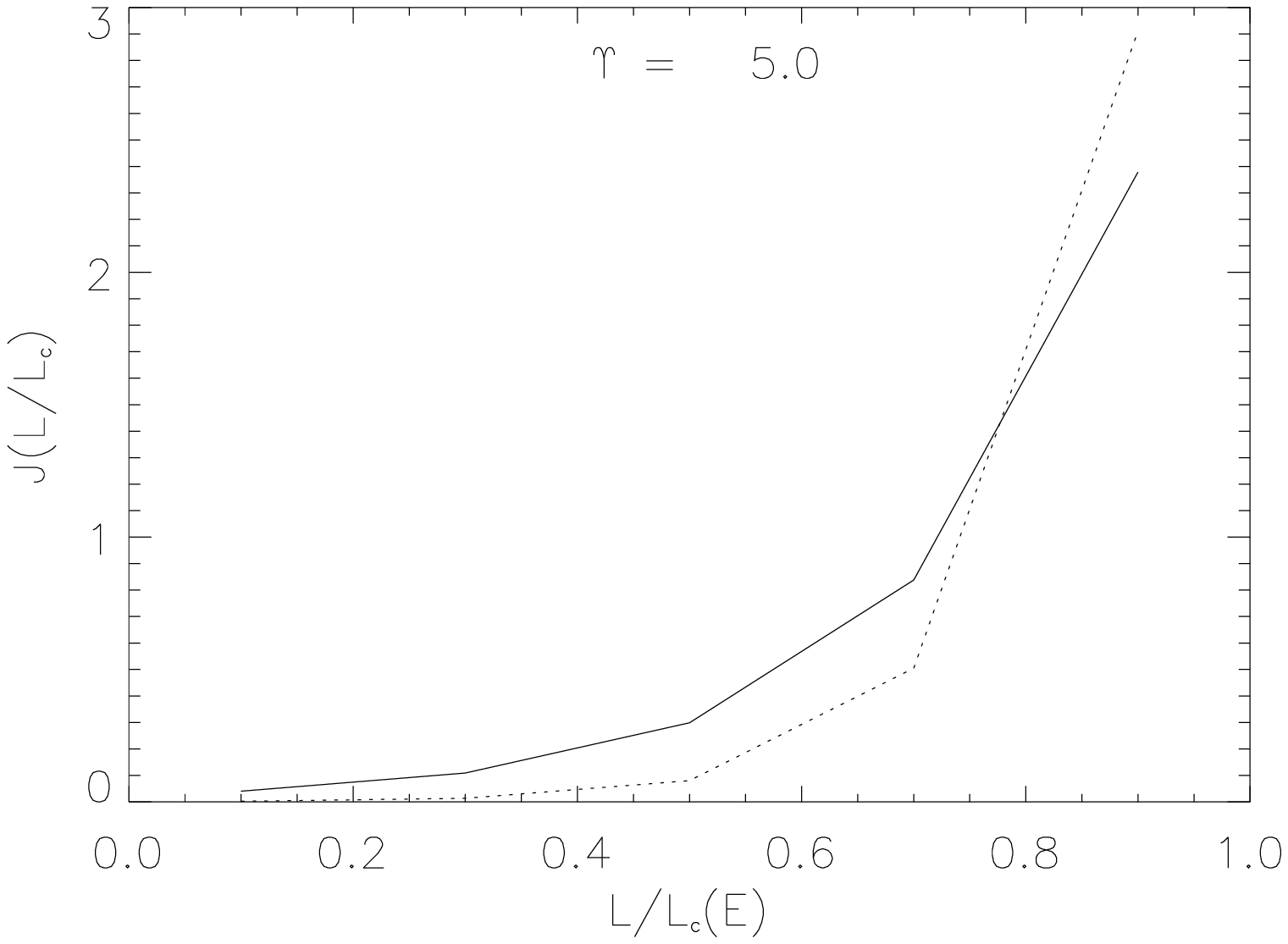}
\epsscale{1.0}
\figcaption{Indicators of anisotropy $J(L/L_c)$ (eq. \ref{eq:JL}). The solid lines
show the input and the dotted 
lines show the derived anisotropy indicators associated with the best-fit models in Figure
\ref{fig:fig1}. For the tangential ($\gamma=5$)
DF, we recover the anisotropy correctly,
but not for the radial DF ($\gamma=-5$). Some simulations with an isotropic DF 
($\gamma=0$, dashed lines) also fail to recover the correct anisotropy.
Therefore, we can not get a reliable result for anisotropy
from a data set of this kind.
\label{fig:fig3}}
\end{figure}

To determine whether our method can recover the anisotropy of the DF from such a small data set, we 
define an indicator
\begin{eqnarray}
I(E, L) \equiv\frac{f(E, L)}{\bar{f}(E)} \;\; \hbox{or} \;\; I_{mn}\equiv\frac{f_{mn}}{\bar{f}(E)}, \label{eq:IEL}
\end{eqnarray}
where $\bar{f}(E)$ is the DF averaged over phase space at energy $E$,
\begin{eqnarray}
\bar{f}(E) &\equiv& \frac{\int_{E}^{E+\delta E} f(E, L) d^3\ve{x}d^3\ve{v}}{\int_{E}^{E+\delta E} d^3\ve{x}d^3\ve{v}} \nonumber \\ 
           &= &{\frac{\sum_{n}w_{mn}}{\sum_{n}V_{mn}}}.
\end{eqnarray}
Figure \ref{fig:fig3} shows the weighted indicator of anisotropy
\begin{eqnarray}
J(L/L_c) \equiv J_n \equiv \sum_{m}I_{mn}U_m,  \label{eq:JL}
\end{eqnarray}
which is a constant for an isotropic DF. The solid lines
show the input and the dotted
lines show the derived anisotropy indicators associated with the best-fit models in Figure
\ref{fig:fig1}. For the isotropic ($\gamma=0$, dotted line) and tangential ($\gamma=5$)
DFs, we recover the anisotropy correctly,
but not for the radial DF ($\gamma=-5$) and some simulations with an isotropic DF              
($\gamma=0$, dashed line).
Our simulations show that there is a high 
probability (5 out of 20 runs) of failing to recover the anisotropy correctly.
Therefore, we can not get a reliable result for anisotropy
from a small ($\sim$160 velocities) data set with these parameters.

Our simulations suggest that we can recover the dependence of the DF on energy $E$ reasonably well with
a sample of this size but
we can not reliably recover the dependence of DF on angular momentum $L$. This is probably 
because the observable distribution 
$g(R, v_z)$ depends more strongly on how GCs are distributed in energy than angular momentum.
Of course, with more data, we may constrain the anisotropy of the DF more tightly.

\section{Mass distribution of M87}
\label{sec:M87}
\subsection{A uniform dataset}
\label{sec:M87data}
\begin{figure}
\plotone{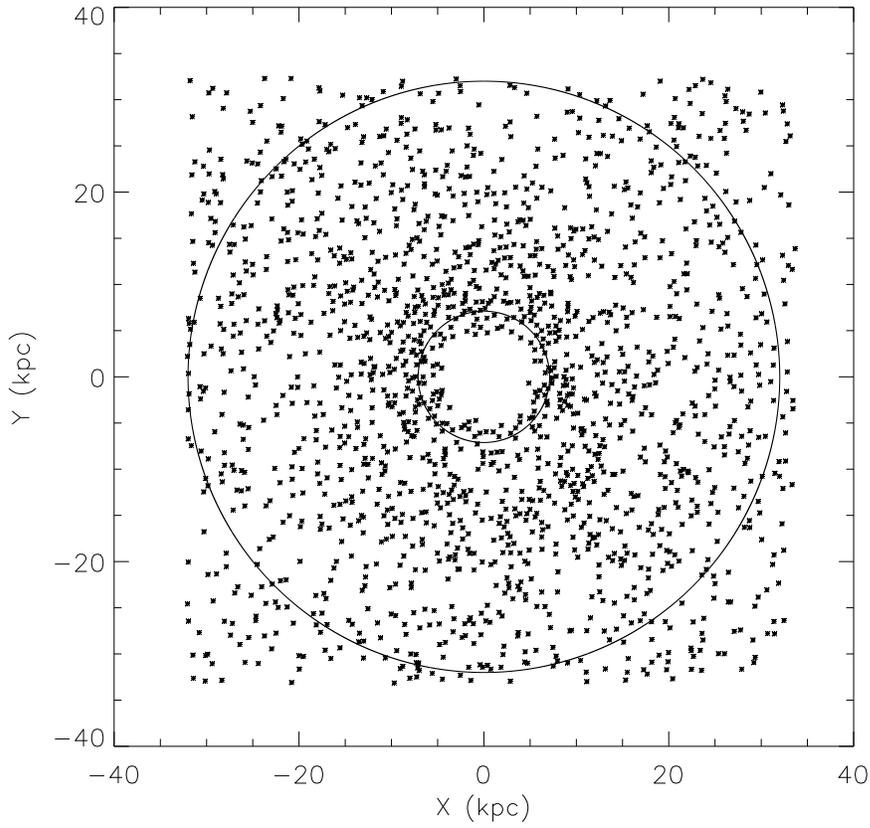}
\figcaption{Spatial distribution of GC candidates in \citet{Coh97}, which are taken from the photometric survey by \citet{Str81}. 
To eliminate selection effects, we only use data between the two circles,
which are 90\arcsec--405\arcsec  or 7--32 kpc from the center, where the Strom et al. survey was complete. 
\label{fig:fig4}}
\end{figure}

\begin{figure}
\plotone{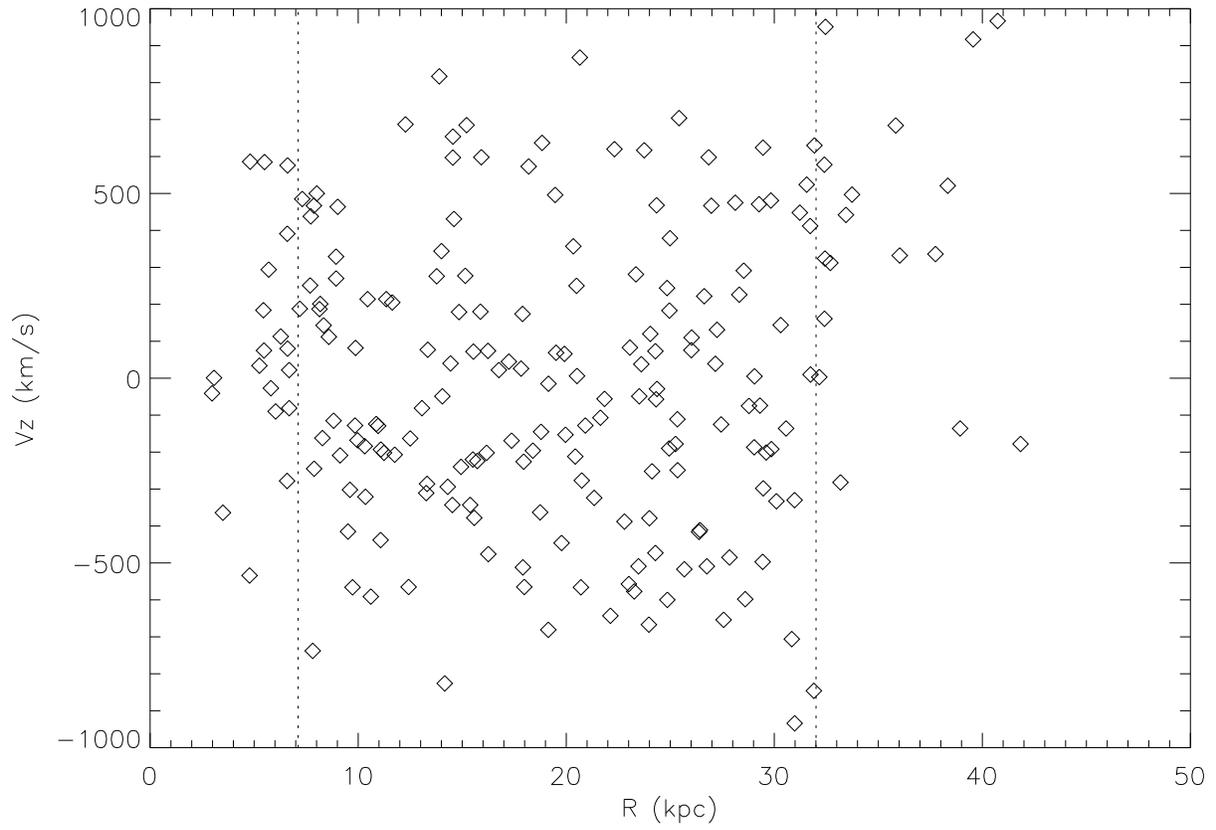}
\figcaption{The data pairs $[R_i, v_{zi}]$ from \citet{Coh97} that we use to construct a uniform dataset.
The data outside the survey radii (dotted lines) are excluded. 
\label{fig:fig5}}
\end{figure}

Surveys of GC positions and radial velocities for M87 are described in Section \ref{sec:intro}. The largest
survey of GC radial velocities is given by \citet{Cot01}, who list 278 GCs. However,
they are drawn from a variety of sources. To construct a uniform sample, 
we use only the data in \citet{Coh97}: these are drawn from a survey 
by \citet{Str81}, which was complete only in the radius range 
$90\arcsec$--$405\arcsec$. So we have the survey limits $(R_{s0}, R_{s1})=(7,32)$ kpc (see Figure \ref{fig:fig4}). Restricting our sample to this range,
we have 161 GCs, which are shown in Figure \ref{fig:fig5}.
The errors in radial velocities given by \citet{Coh97} range between 50--100 km s$^{-1}$; we set 
the errors equal to 75 km s$^{-1}$ for all GCs.

   We also incorporate the surface number density profile $\Sigma_{obs}(R_j)$, which
is given with estimated error $\sigma_\Sigma(R_j)$
 in 25 bins from
7 to 110 kpc by \citet{Har86} and \citet{Mcl99}, as an additional constraint in our calculation, 
as described in 
\S\ref{sec:tec} . 
Both $\Sigma(R)$ and $\Sigma_{obs}(R)$ should be normalized so that $2\pi\int RdR\Sigma(R)$ and
$2\pi\int RdR\Sigma_{obs}(R)$ are unity within the radius range 7--110 kpc. Notice that we do double count
some of the GCs in 7--32 kpc since they appear in both the kinematic data $[R_i, v_{zi}]$ and 
in the surface density distribution $\Sigma_0(R_j)$.
However, this does not affect our results in a noticeable way because there
are over 7000 GCs used in the calculation of the surface density distribution.

\subsection{Analysis}

   We have described our method in \S\ref{sec:tec}. In constructing our models we must
choose a maximum energy $E_{N_E}$ (eq. \ref{eq:EL}) for the cluster population, or equivalently, a largest apocenter
$r_{m}$ such that $\Phi(r_m)=E_{N_E}$. We choose $r_m=300$ kpc, 9 times larger than the outer survey
radius $R_{s1}=32$ kpc. If $r_m$ is too small, we bias our results by excluding high-energy
GCs that may be present within the survey radii. If $r_m$ is too large, the numerical work will
be dominated by GCs that spend only a small fraction of their orbits within the survey radii. For comparison, 
\citet{Har86} observed GCs out to projected radius $R=110$ kpc.

\subsection{Results}
\subsubsection{Isotropic power-law models}
\label{sec:iso}

   Initially we assume 
that the DF is isotropic ($N_E=80$, $N_L=1$) and the mass density profile of the galaxy is a power law (eq. \ref{eq:powerlaw}). We
set the arbitrary radius ${r_0}$ in equation (\ref{eq:powerlaw}) to be 19 kpc, near the median radius of our
samples, to minimize the covariance of $\rho_0$ and $\alpha$
in their bivariate probability distribution. 

\begin{figure}
\plottwo{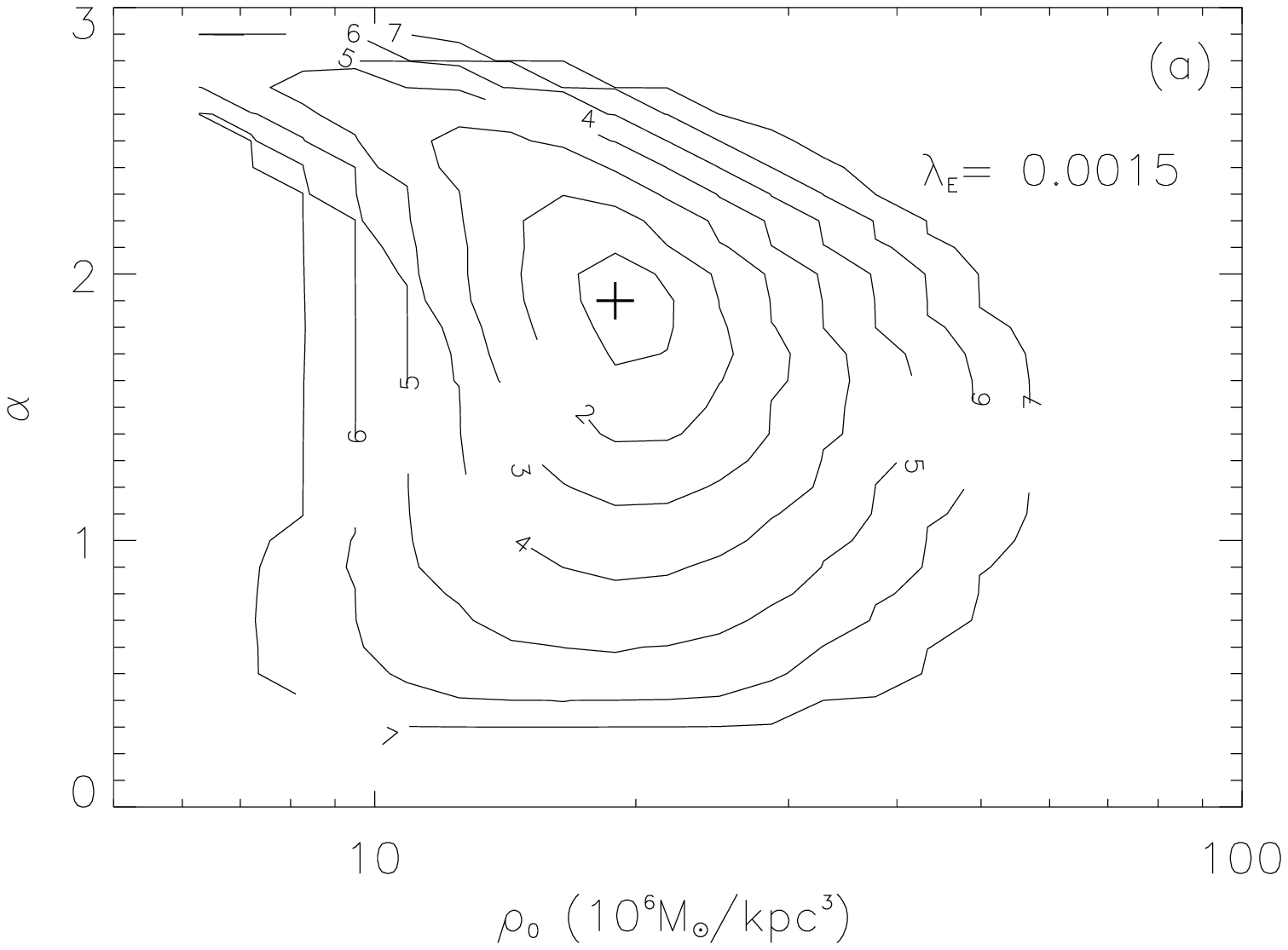}{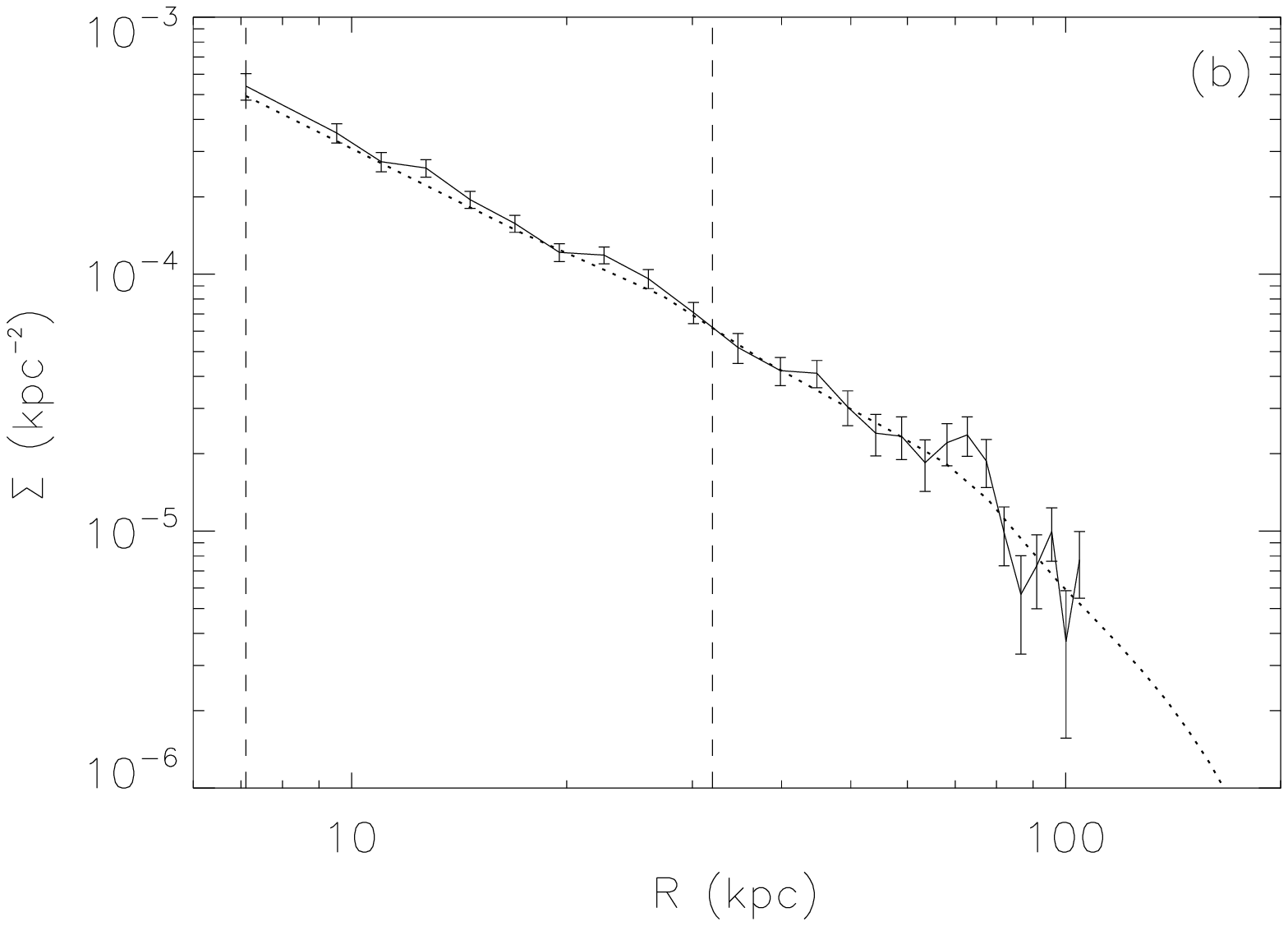}
\figcaption{Fitting the kinematic data $[R_i, v_{zi}]$ and the observed surface number density $\Sigma_{obs}(R)$ of GCs in M87 to a power-law 
potential, assuming an isotropic DF. (a) Contour plot of $Q'_{max}(\ve{X})$ as a function of $\ve{X}=\{\rho_0, \alpha\}$.
(b) The derived surface number
density (dotted line) from maximizing $Q'(\ve{X}, \ve{W})$, compared to the observed $\Sigma_0(R)$ (solid line). 
The surface densities are
normalized so that the total number within the range 7--110 kpc is unity. The kinematic survey
limits are marked by dashed lines.
\label{fig:fig6}}
\end{figure}

\begin{figure}
\plotone{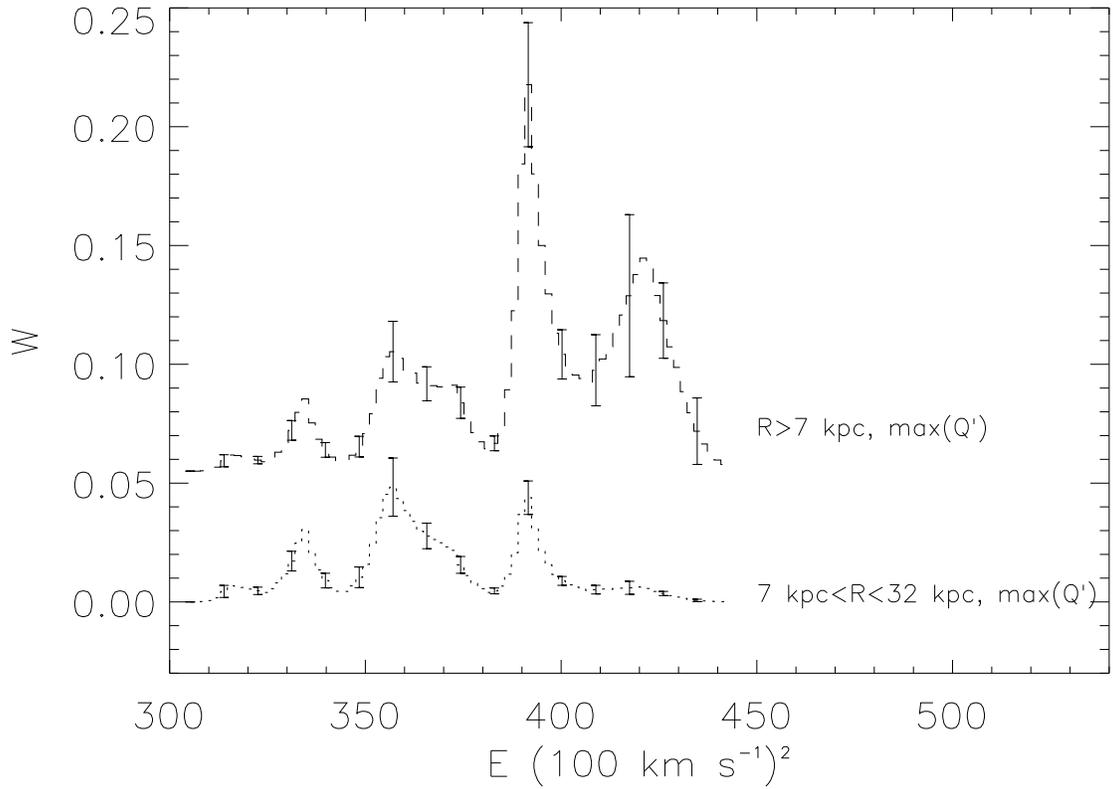}
\figcaption{
Energy distribution with the isotropic assumption: the energy weights $\ve{W}$ as determined by fitting
both the kinematic data and the surface density profile, as in Figure \ref{fig:fig6}. 
The bottom dotted line refers to the energy distribution of GCs
within projected radius $R=7-32$ kpc and the upper dashed line to all GCs with $R\geq 7$ kpc.
The error bars are estimated from bootstrap resampling.
\label{fig:fig7}}
\end{figure}

In this case, $\ve{X}=\{\rho_0, \alpha\}$. Figures \ref{fig:fig6}a,b show 
the contour plot of $Q'_{max}(\ve{X})$ and the resulting best fit to the surface number density profile.
The appropriate smoothing parameters are $(\lambda_E, \lambda_L)=(0.015, 0.015)$ so that $n_s=2$ as
described in \S\ref{sec:tec}.
We get a
very tight constraint on the potential if the DF is assumed isotropic.
The isotropic models give $M($32 kpc$)=(2.6\pm0.3)\times 10^{12}M{_\odot}$ 
and $\alpha=1.8\pm0.2$.
The best fit value of $\alpha$ is consistent with
the value 1.7 found by \citet{Rom01} for similar isotropic models.
Figure \ref{fig:fig7} shows the derived energy distribution of the GCs. 
The dotted line (bottom) 
is the energy distribution of GCs within projected radius $R=7-32$ kpc, which
is calculated from the DF associated with the best-fit model in Figure \ref{fig:fig6}a.
The upper dashed line is the energy distribution of all GCs with
$R\geq7$ kpc in the same model.
The error bars are estimated from bootstrap resampling.
The energy distributions suggest that there may be kinematically distinct groups of GCs in M87.

\subsubsection{Anisotropic power-law models}

\begin{figure}
\plottwo{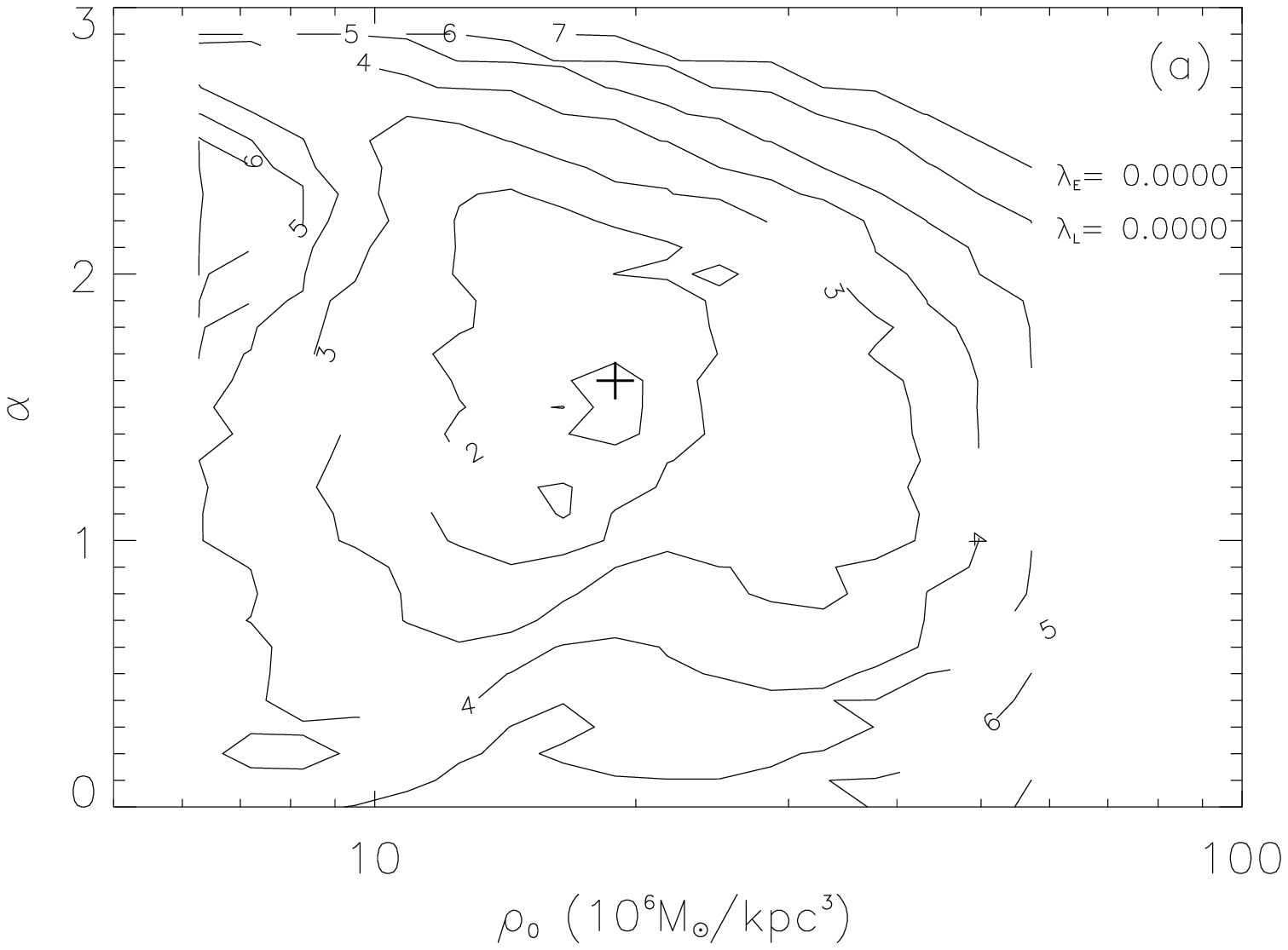}{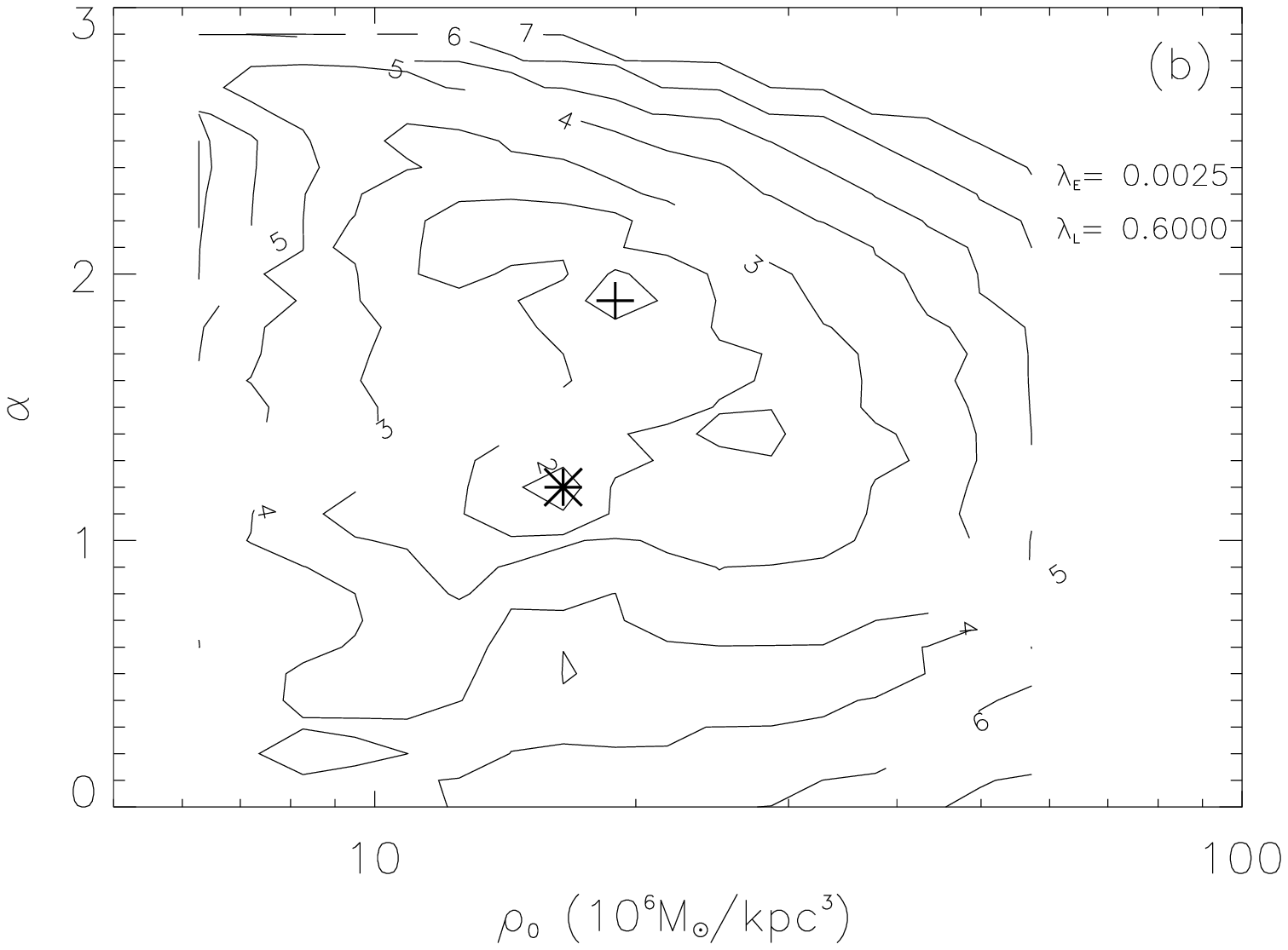}
\plottwo{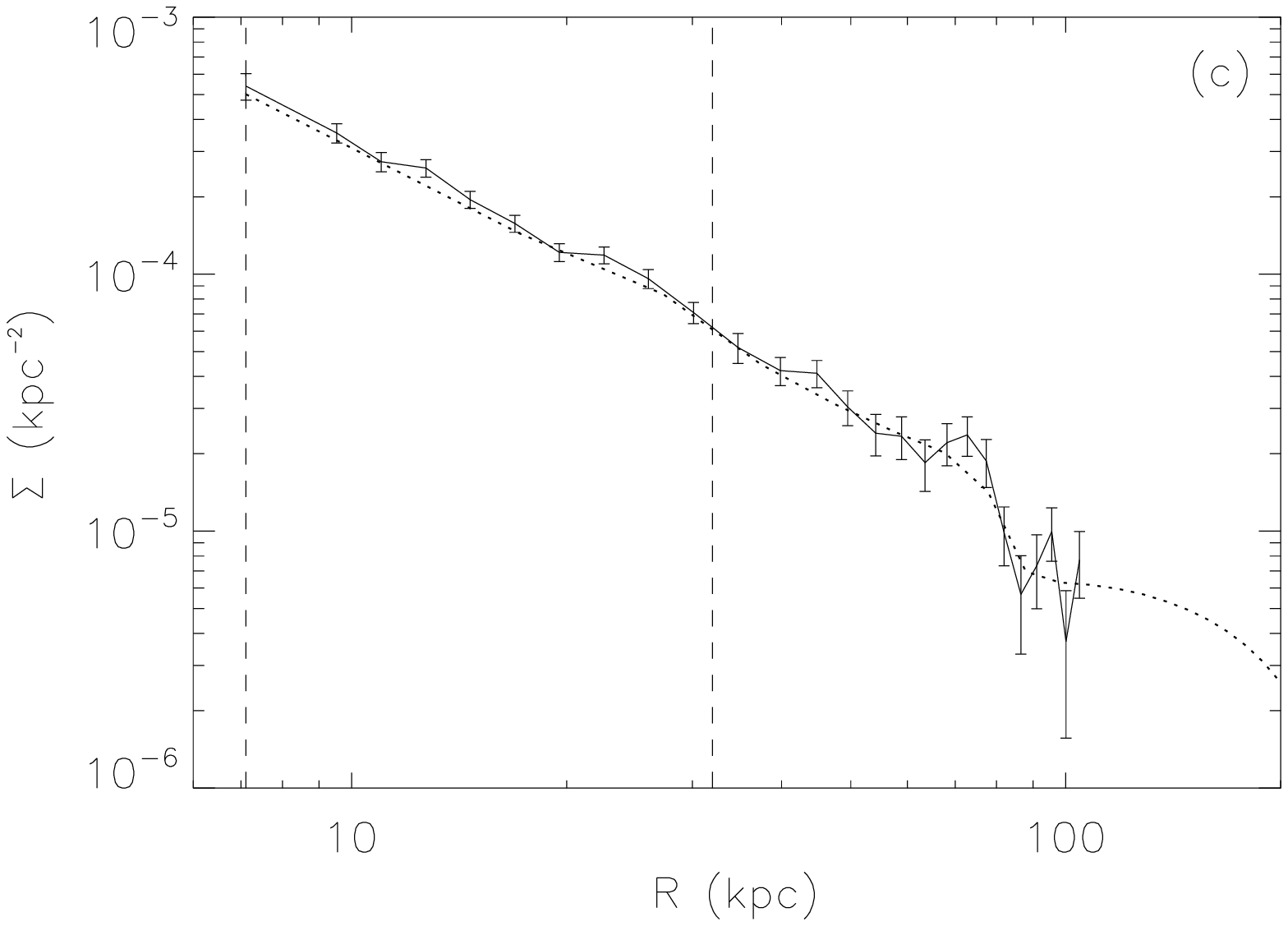}{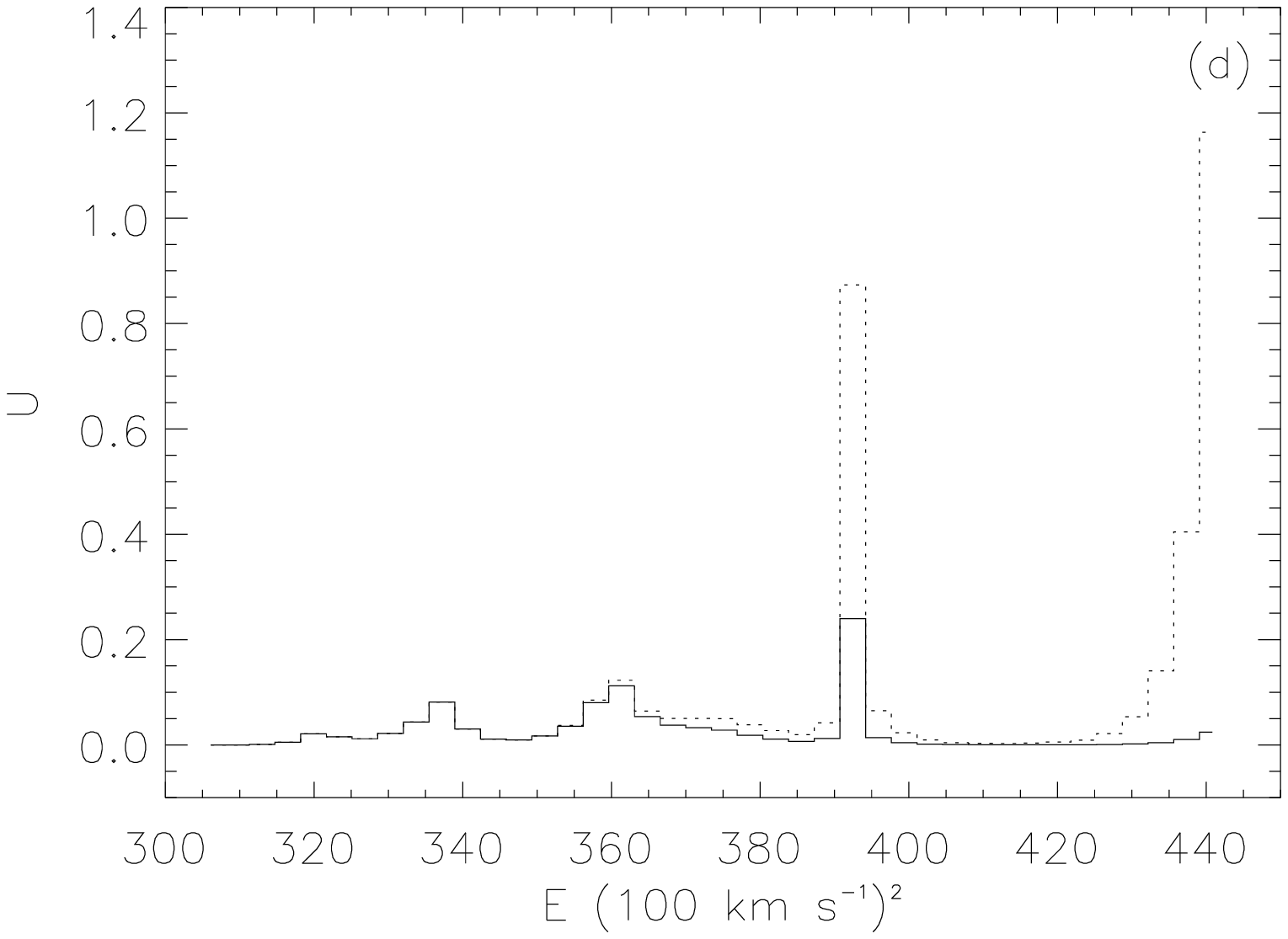}
\figcaption{Fitting the kinematic data $[R_i, v_{zi}]$ and the observed surface number density $\Sigma_{obs}(R)$ to a power-law
potential without assuming isotropy of the DF. (a) The same as Figure \ref{fig:fig6}a except for anisotropic models without regularization.
(b) The same as Figure \ref{fig:fig8}a except for anisotropic models with appropriate smoothing parameters.
The plus and asterisk mark the best-fit model and a secondary peak, respectively.
(c) The same as Figure \ref{fig:fig6}b except for the best-fit anisotropic model with appropriate smoothing parameters.
(d) Energy distribution of GCs 
within projected radius $R=7-32$ kpc (solid line) and with $R\geq 7$ kpc (dotted line).
\label{fig:fig8}}
\end{figure}

Figures \ref{fig:fig8}a and \ref{fig:fig8}b show the result of fitting models with power-law potentials and
anisotropic DFs ($N_E=40$, $N_L=5$) to both the kinematic data and
the observed surface number density, using both $\lambda_E, \lambda_L=0$ and the
appropriate smoothing parameters $(\lambda_E, \lambda_L)=(0.0025, 0.60)$ so that $n_s=2$, respectively.
The contours in Figure \ref{fig:fig8}a (without regularization) and
Figure \ref{fig:fig8}b (with appropriate smoothing parameters)
are very similar, suggesting that the estimate of potential parameters of M87 is rather insensitive to 
smoothing parameters.
The contour in Figure \ref{fig:fig8}b is 
more irregular and the uncertainties are larger than in Figure \ref{fig:fig6}a, 
which is expected since anisotropy gives
more freedom to adjust the DF to fit the data; however, the best-fit potential parameters $(\rho_0, \alpha)
=(1.9\times10^7\,\hbox{M}_{\sun}\,\hbox{kpc}^{-3}, 1.9)$ (plus symbol) do not change from those for isotropic models. 
A secondary peak, marked by the asterisk symbol, has $Q'$ smaller by 0.2.
The derived surface number density from the best-fit model in Figure \ref{fig:fig8}c
has a stronger tail than that derived under the assumption of isotropy (see Figure \ref{fig:fig6}b).
Figure \ref{fig:fig8}d
shows the energy distributions of GCs
within projected radius $R=7-32$ kpc (solid line) and with $R\geq 7$ kpc (dotted line), which
are also concentrated into a few peaks. The peak at $E\simeq440\times(100\,\hbox{km s}^{-1})^2$ accounts for the strong tail in the derived 
surface number density profile beyond 100 kpc. 

These results are based on the estimate of background stars $5.8\pm0.3$ stars arcmin$^{-2}$ \citep{Har86}. 
We also investigate the case of $6.1$ stars arcmin$^{-2}$. We get almost the same
likelihood contours and the best-fit potential parameters change by less than $1-\sigma$; 
however, the amplitude of the high-energy tail is
reduced by a factor of 2.

\begin{figure}
\plottwo{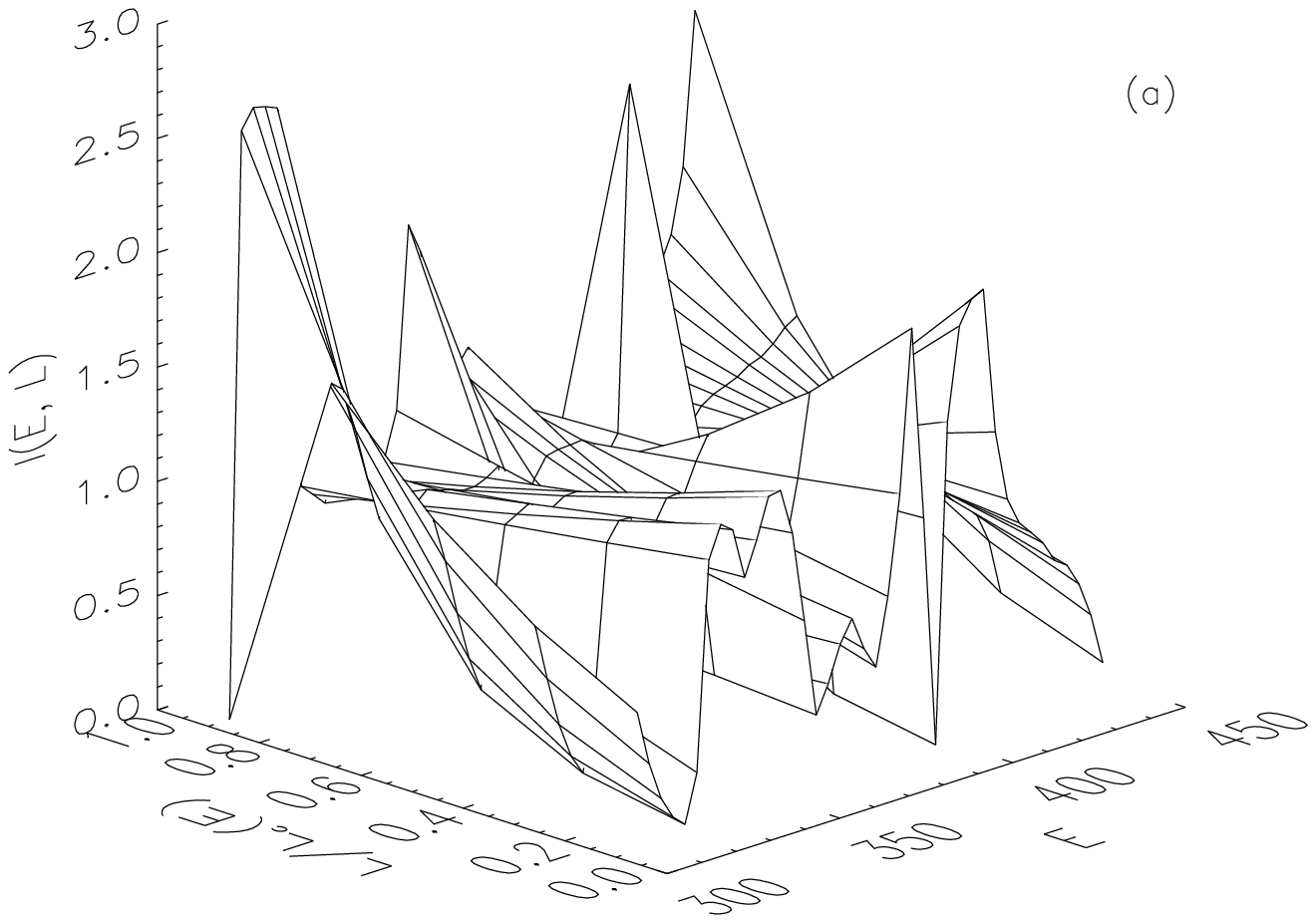}{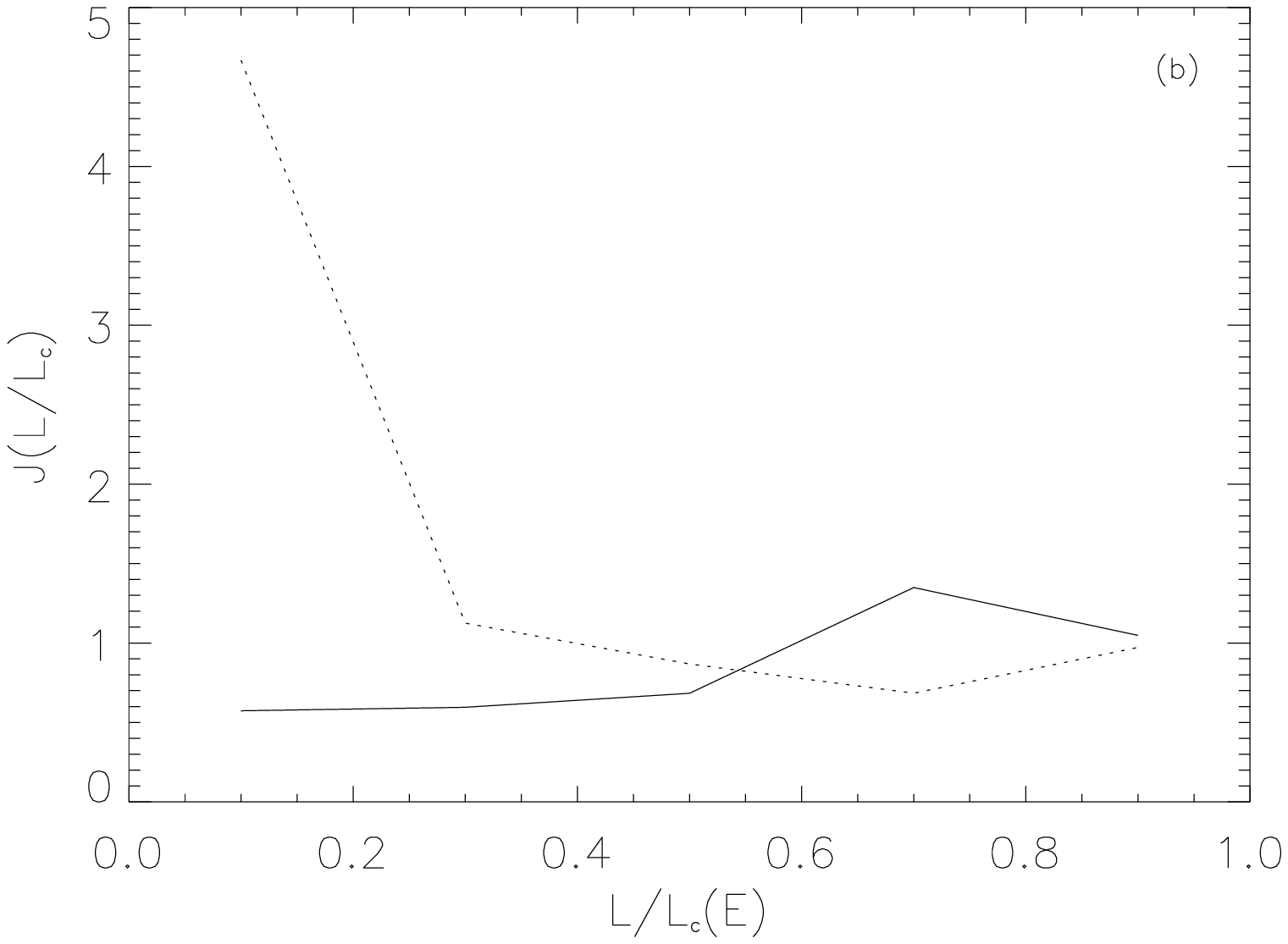}
\figcaption{Indicators of anisotropy for $\lambda_E, \lambda_L=0.005$. (a) $I(E,L)$ (eq. \ref{eq:IEL}). 
(b) $J(L/L_c)$ (eq. \ref{eq:JL}) for the best-fit model (solid line) and at the secondary peak (dotted line) in Figure \ref{fig:fig8}b. The behavior of the indicators suggests that 
the GCs may either prefer circular orbits (high angular momentum) or radial orbits (low angular momentum),
depending on the mass distribution model of the galaxy.
\label{fig:fig9}}
\end{figure}

Figure \ref{fig:fig9} shows the indicators of anisotropy $I(E,L)$ and $J(L/L_c)$. The results suggest that
the GCs prefer circular orbits (high angular momentum) for the best-fit model but
radial orbits (low angular momentum) at the secondary peak.
To reliably determine the anisotropy of the GCs, we need to obtain more data to give a tighter
constraint on mass distribution and develop a more robust formalism of choosing smoothing
parameters in the future.

\begin{figure}
\plottwo{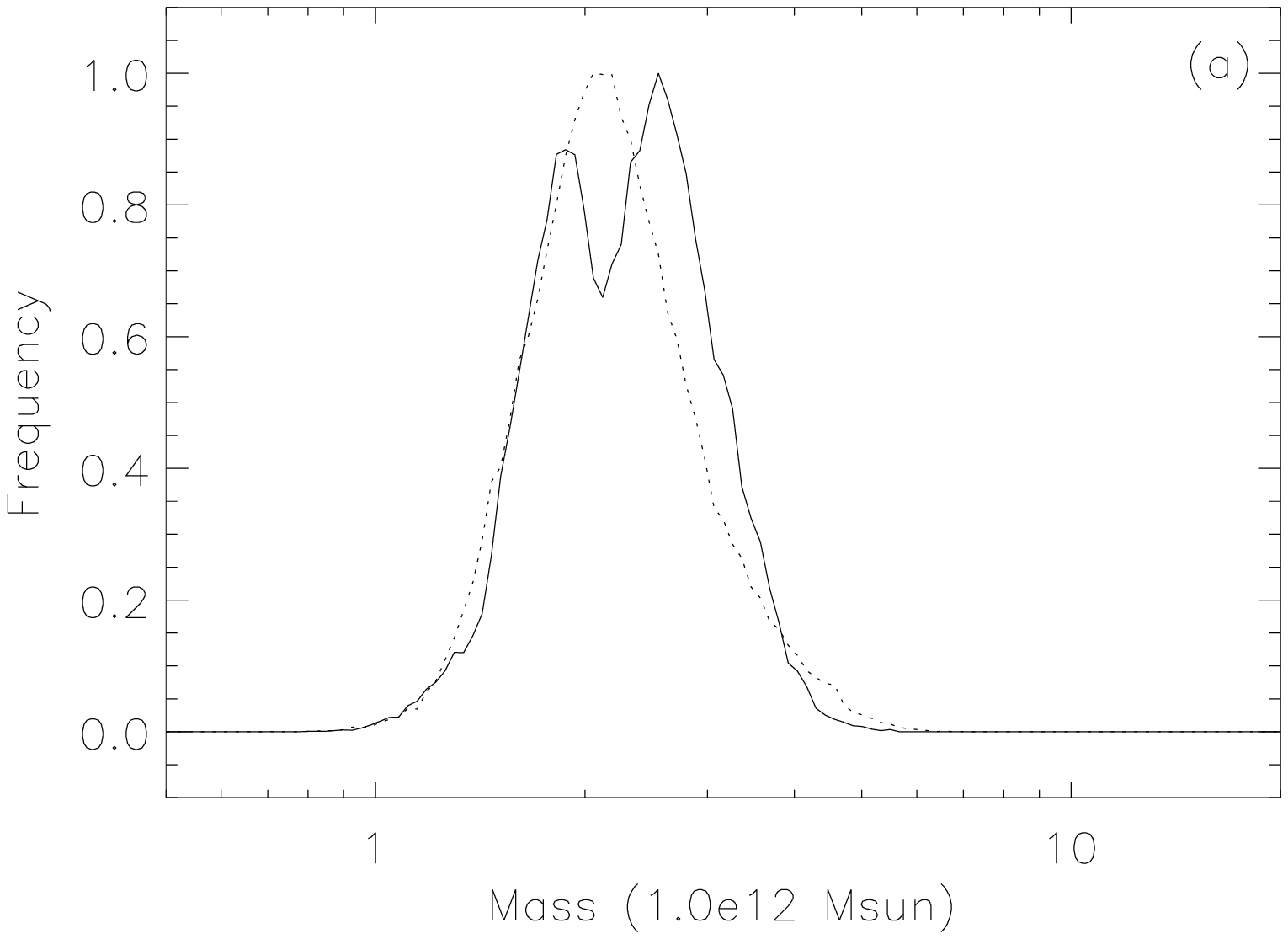}{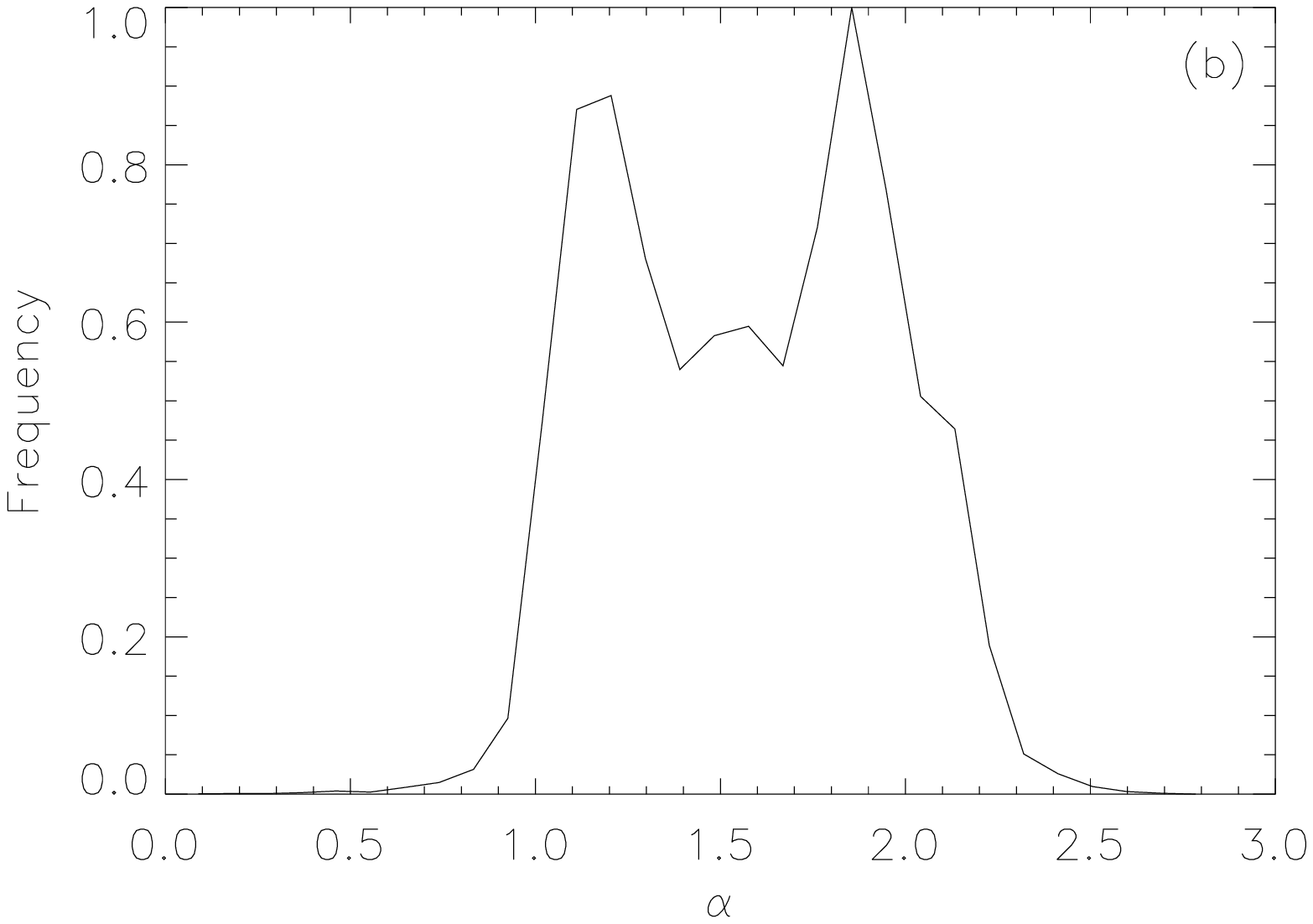}
\figcaption{Distribution of $M(r=32$ kpc$)$ and $\alpha$ for likelihood distribution of Figure \ref{fig:fig8}b. 
(a) The probability
distribution of $M(r=$32 kpc$)$ for the power-law (solid line) and NFW models (dotted line). 
(b) Probability function of $\alpha$ for the power-law models.  It gives $\alpha=1.6\pm0.4$.
\label{fig:fig10}}
\end{figure}

To estimate the probability distribution of parameters derived from the potential,
we lay down points
in Figure \ref{fig:fig8}b with a uniform probability distribution in $(\ln\rho_0, \alpha)$, then reject or save each point according to $Q'(\ve{X}, \ve{W})$ at that point.
From the saved
points, we may estimate the distribution of $M(r)$, the mass enclosed within radius $r$, and the exponent 
$\alpha$.
Figure \ref{fig:fig10}a shows the
probability distribution of $M(r=32$ kpc$)$, which is derived from $4\times10^4$ points and has been normalized
to its maximum. 
Figure \ref{fig:fig10}b shows the probability distribution
of the density slope $\alpha$. 
The data provide a fairly tight constraint on $\alpha$, which is $1.6\pm0.4$ (1--$\sigma$ error).

\begin{figure}
\plottwo{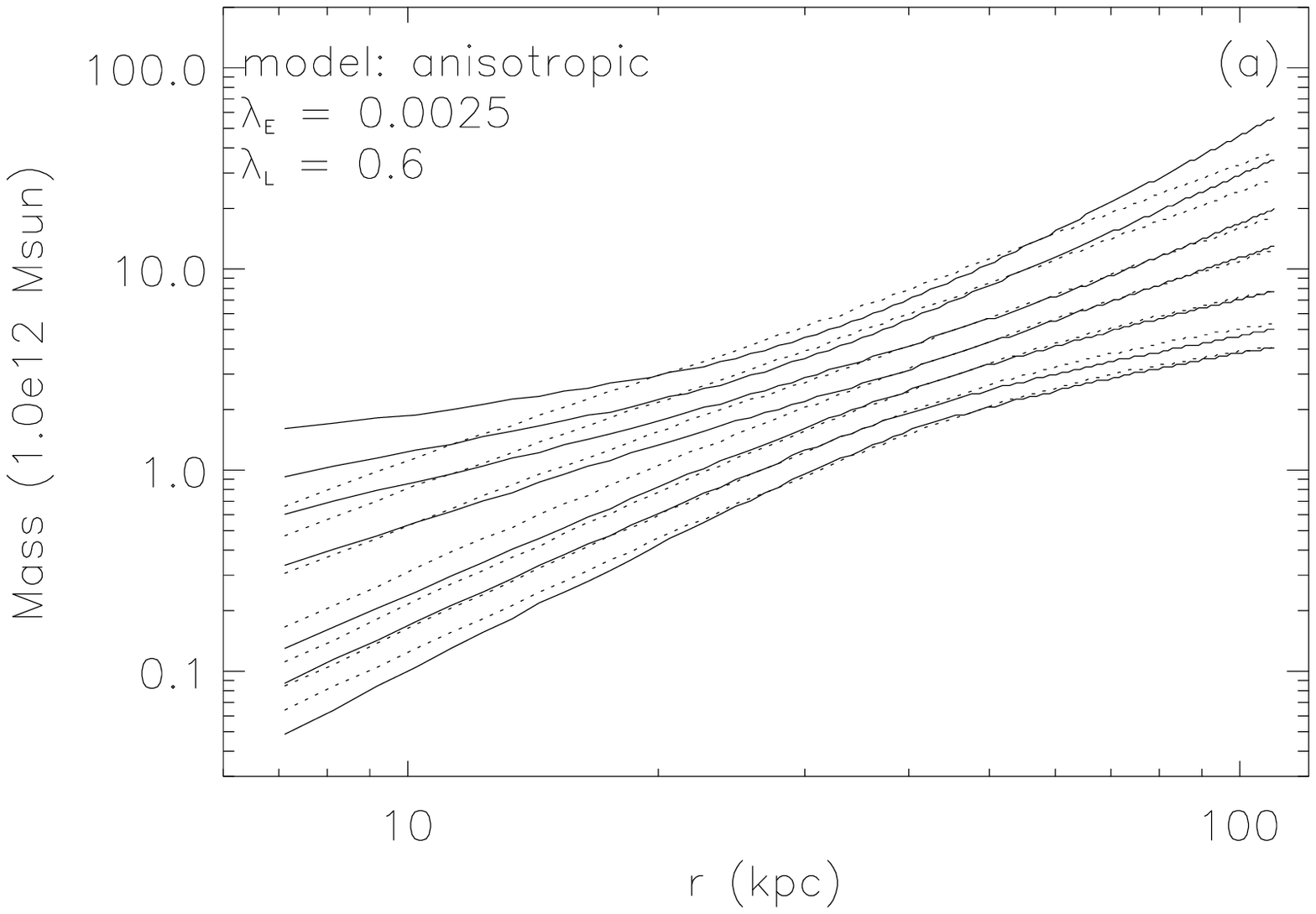}{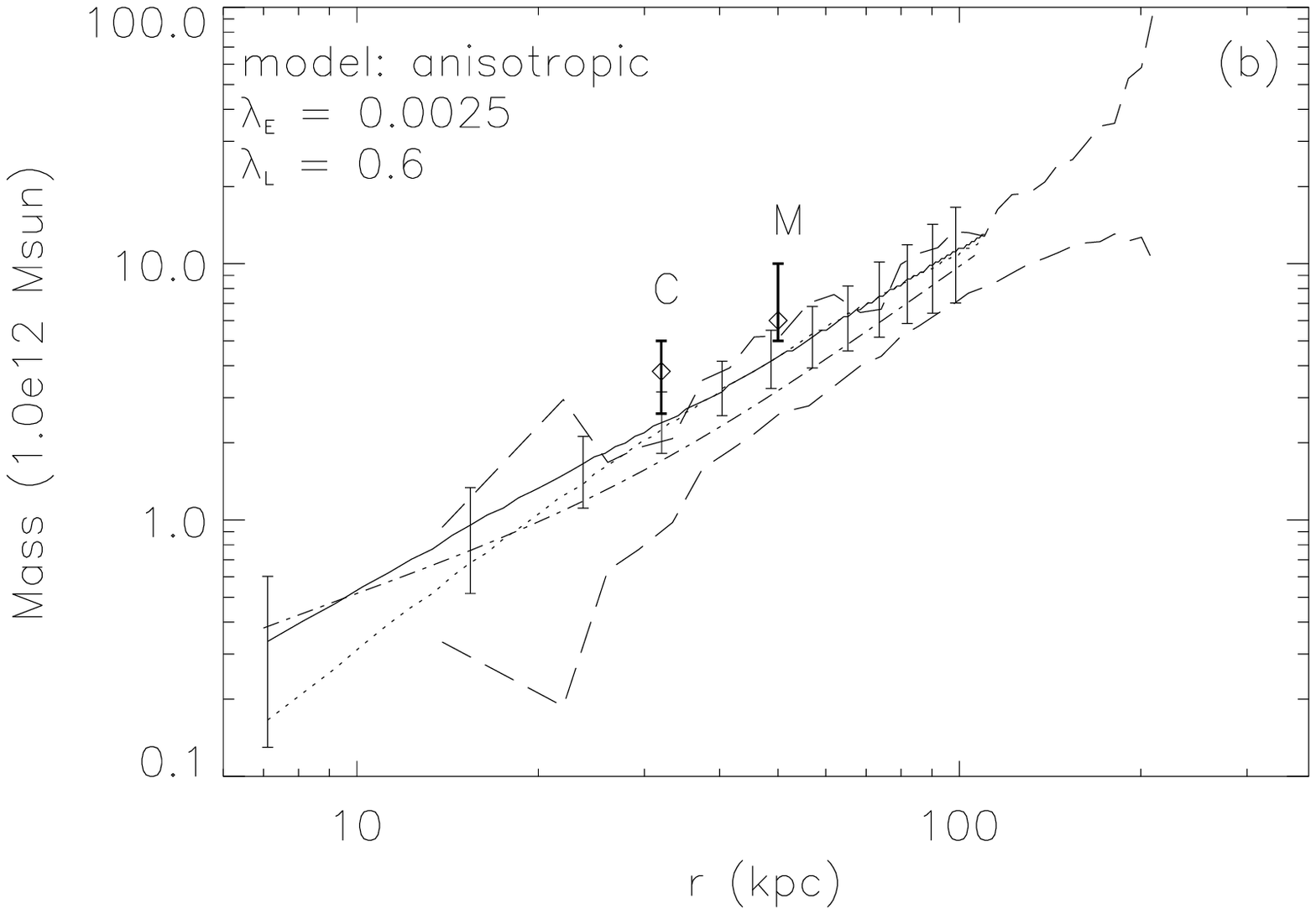}
\figcaption{Estimates of $M(r)$, derived from Monte Carlo points laid down on 
Figure \ref{fig:fig8}b. 
(a) The probability distribution of $M(r)$. The lines
show the median and 1,2,3--$\sigma$ errors which
enclose the regions with the cumulative probabilities of 68.3$\%$, 95.4$\%$, 99.7$\%$ for the
power-law models (solid lines) and the NFW models (dotted lines).
(b) The solid and dotted lines 
show the estimates for $M(r)$ by
assuming power-law and NFW density profiles, respectively. 
The error bars indicate the uncertainties in the mass estimates, which are similar for
both models.
The other models are: \citet{MT93} (marked by ``M''), 
\citet{Coh97} (marked by ``C''), upper and lower limits from X-ray observation by \citet{Nul95} (long dashed 
lines), \citet{Mcl99b} (dot-dashed line). 
The best-fit model NFW2 (star+halo) in \citet{Rom01}
is within 5\% of our best-fit power-law model from 7 kpc to 110 kpc, thus is not shown for clarity. However,
considering the large error bars ($>30\%$) in our models, it is probably a coincidence that
the two results are so close. Our result is also consistent with X-ray observation by \citet{Mat02} within
20\% from 15 to 80 kpc by reading numbers from the double $\beta$ model fit in their Figure 21.
\label{fig:fig11}}
\end{figure}

If we stack curves like that in Figure \ref{fig:fig10}a for different radii, we get the
mass distribution for all radii, which is shown in Figure \ref{fig:fig11}a. The solid curves mark
the mean and 1,2,3-$\sigma$ errors. 
The solid line and error bars in Figure \ref{fig:fig11}b show the best estimate of $M(r)$ and its
standard deviation,
which gives $M($32 kpc$)=(2.4\pm0.6)\times 10^{12}\,\hbox{M}{_\odot}$, 
differing from that in isotropic models by less than the error bars.

Notice that the
relative error is the smallest around 32 kpc, which is reasonable since it is the maximum survey radius. 
The mass profile can be approximately described as an analytical form,
\begin{eqnarray}
M(r)=(2.3r^{1.36})_{-(55-1.9r+0.056r^2)} ^{+(63-2.4r+0.071r^2)}\times10^{10}\,\hbox{M}_{\odot}, \;\; 
15\;\hbox{kpc}<r<110\;\hbox{kpc},
\end{eqnarray}
where $r$ is in units of kpc.

Also shown in Figure \ref{fig:fig11}b are previously published models. The data for X-ray
observations \citep{Nul95} were obtained by assuming a distance of 20 Mpc to M87. 
The long
dashed lines show
their derived lower and higher limits scaled by a factor of 0.815 since we assume a distance of 16.3 Mpc. 
Our mass estimate is consistent with the X-ray observations in the range where they overlap.
Our result is also consistent with X-ray observation by \citet{Mat02} within
20\% from 15 to 80 kpc by reading numbers from the double $\beta$ model fit in their Figure 21.
The estimate by \citet{Mcl99b} is based on X-ray observations and an assumption of
a NFW dark matter halo of the Virgo cluster.
The estimates
by \citet{Mer93} and \citet{Coh97} are based on the assumption of an isotropic DF.

\begin{figure}
\plotone{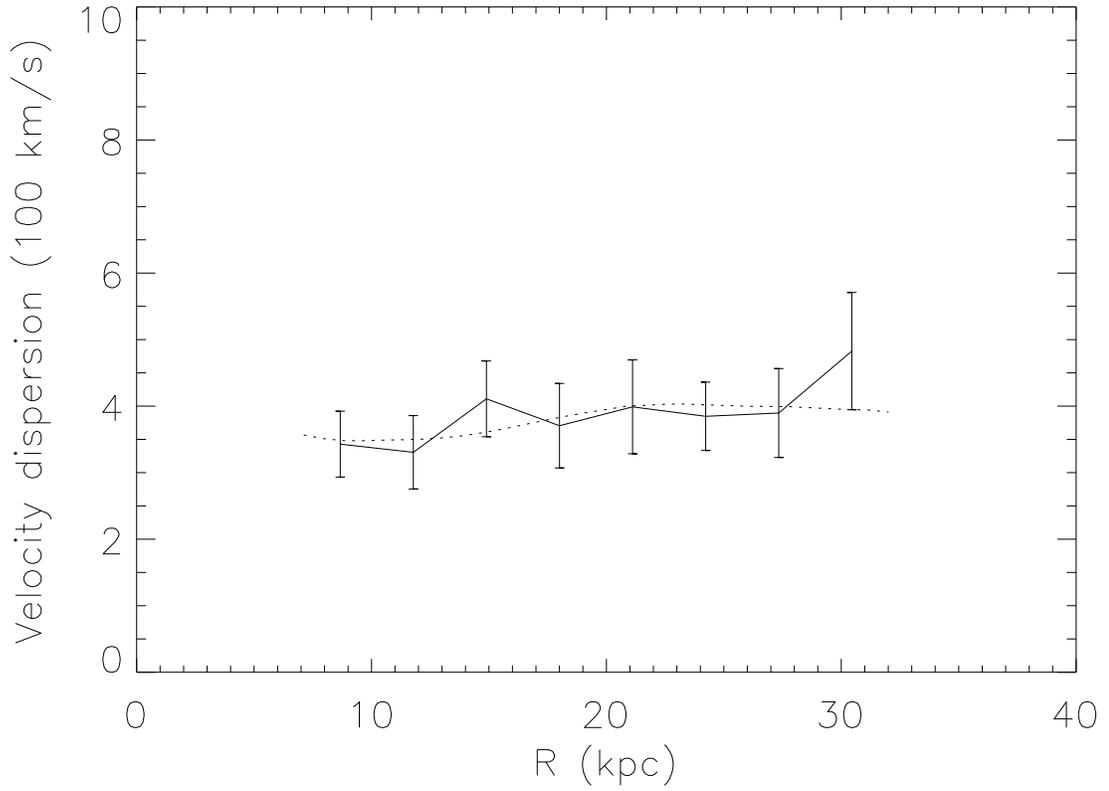}
\figcaption{Velocity dispersion, including assumed observational errors of 75 km$\,$s$^{-1}$
in the theoretical model. The dotted line is derived from the best-fit anisotropic power-law model, which is 
consistent with the observations (solid line with error bars).
\label{fig:fig12}}
\end{figure}

Figure \ref{fig:fig12} shows the derived velocity dispersion profile (dotted line)
\begin{eqnarray}
\sigma_z(R)={\langle v_z^2 \rangle ^{1/2}}=\left (\frac{\int v_z^2 g(R, v_z) dv_z}{\int g(R, v_z) dv_z} \right )^{1/2}, 
\end{eqnarray}
which is consistent with the observed velocity dispersion profile (solid
line with error bars).

\subsubsection{Other models}

  We also investigate models with two other potentials, (i)
a NFW profile (eq. \ref{eq:NFW}), (ii) the potential generated by
the stars, assuming constant mass-to-light ratio.
We adopt the B-band luminosity profile in \citet{Mcl99b}, which is smaller
than the luminosity profile in \citet{Rom01} by 23\%, probably due to a
systematic shift between different data they use.

The dotted line in Figure \ref{fig:fig11}b shows the estimate for $M(r)$ for anisotropic 
NFW models,
which is almost the same as that for power-law models beyond 20 kpc.
The errors are comparable
to those for power-law models. 
We explore NFW models with the concentration radius $r_c$ ranging from 10 to 500 kpc.
If the DF is assumed isotropic, we get $r_c=26_{-4}^{+27}$ kpc, while
models with $r_c = 100$ and 500 kpc are ruled out
at 2 and $3-\sigma$ level, respectively. 
However, the best-fit anisotropic NFW models with 10 kpc $<r_c<500$ kpc give $Q'_{max}$ within
$2-\sigma$, i.e., a wide range of $r_c$ can fit the data pretty well. The models NFW1, NFW2 and NFW3 of \citet{Rom01}
are within 1-$\sigma$ relative to our own best fit.

The constant mass-to-light ratio model gives $(M/L)_B=125\pm10\,\hbox{M}_{\sun}/\hbox{L}_{\sun}$,
which is much larger than the value $43\pm3\,\hbox{M}_{\sun}/\hbox{L}_{\sun}$ obtained by \citet{Rom01} fitting the globular cluster kinematics.
The difference in the B-band luminosity profiles can not account for the large gap.
For the same regularization level, the maximum penalized likelihood for the best-fit powerlaw and NFW models differ by less than 1-$\sigma$ while
the maximum for constant $M/L$ is smaller by 7, i.e., the constant $M/L$ model is about 4-$\sigma$ worse than
for the NFW or power-law models.

\section{Discussion}
\label{sec:dis}

The mass distribution of M87 from GCs has been studied by \citet{MT93, Coh97, Rom01}.
In general our results are consistent with the conclusions of these earlier papers.
Our results are more robust than those by \citet{MT93} and \citet{Coh97} since 
we assume a two-integral DF rather than an isotropic DF.
In some aspects, our analysis is less powerful
than that of \citet{Rom01}, because we do not use the stellar part to constrain
the mass distribution of M87.
Therefore,
it is not surprising that we constrain the mass distribution around the effective
radius ($R_e=7$ kpc) rather poorly (see Figure \ref{fig:fig11}).
As to the analysis of GC kinematics,
our methods and results differ from those in \citet{Rom01} in the following aspects.

\begin{enumerate}
\item Romanowsky \& Kochanek (2001) fit their kinematic data to predictions for
the probability distribution of radial velocities at given radii, rather than
to the joint probability distribution of radius and velocity as we do. This
difference in approaches is necessary because they use an incomplete sample of
GCs, whereas ours is complete for $R\in (R_{s0},R_{s1})=(7,32)$ kpc. Our
approach is statistically more powerful but requires us to use a subset of the
GCs with measured radial velocities.

\item We use regularization rather than entropy to smooth our results. As we have argued
in \S\ref{sec:tec}, entropy smoothing tends to favor high-energy orbits. \citet{Rom01}
use entropy only as a numerical device to accelerate convergence to a final solution,
eventually reducing the entropy contribution to a negligible value by reducing
the constant $\alpha$ in equation (\ref{eq:alpha}) to a very small value. Nevertheless, this
process may still tend to bias the solution towards high-energy orbits.

\item Although \citet{Rom01} explore both the best fit and
the uncertainties in their mass models, they do not rigorously explore the range
of DFs and velocity anisotropies that are consistent with the data.
We find that either radial or tangential anisotropy in the DF are allowed
by the data. We also explore a wider range of mass models.

\end{enumerate}

The observations of metallicity of GCs show that there are metal-rich and
metal-poor globular cluster samples in M87. The metal-rich sample is more concentrated
to the center \citep{Cot01}. We tried to divide all GCs into two samples to
study their kinematics individually. Ideally it would be interesting to see how the subsamples
contribute to the peaks observed in the energy distribution in Figure \ref{fig:fig7} and
\ref{fig:fig8}d. However, the present data set is too small so that the statistics is too
poor to reach a conclusion. 

An important question is whether the mass distribution we have measured
is a dark halo associated with M87 or with the Virgo cluster as whole. If
the DF is isotropic, we find that $r_c=26_{-4}^{+27}$ kpc,
which implies that the halo is small enough that it must be associated with M87. However,
for general DFs all halos with 10 $<r_c < 500$ kpc fit
the data, so we cannot determine whether the halo is associated with
the galaxy or its host cluster.

\section{Conclusions}
\label{sec:con}

We have investigated how to determine the mass distribution of a spherical
stellar system from the kinematics of a discrete set of test particles. We use
a non-parametric form for the DF, which
includes the isotropic DF as a special
case, and parametrized forms for the potential,
and find the best-fit potential and DF using a penalized maximum
likelihood method. 
Our method allows for selection effects, 
observational and statistical errors, and
anisotropy in the DF.
We find that
the potential parameters are determined more securely than the energy
distribution, which is in turn more accurate than the angular-momentum
distribution.
Our simulations show that mass distributions depending on two parameters
can be derived rather accurately
from a small number ($\leq 200$) of data pairs $[R_i, v_{zi}]$ and
surface number density profile.
Plausible estimates of the energy distribution of
the test particles can also be recovered; however, we can not tell whether the DFs are isotropic
or not from such a small dataset.

We have applied our method to a sample of 161 GCs between 7 kpc and 32 kpc in M87.
   Under the assumption that the system is spherical and the DF is isotropic,
we infer that the mass of M87 within 32 kpc is $2.6\pm0.3\times 10^{12}\,$M${_\odot}$. The
power-law index for the density profile, $\rho\propto r^{-\alpha}$ is $\alpha=1.8\pm0.2$. 
The energy distribution suggests a few kinematically
distinct groups of globular clusters.
If we allow an anisotropic DF, the mass of M87 within 32 kpc is $(2.4\pm0.6)\times 10^{12}\,$M${_\odot}$
and the power-law index is $1.6\pm0.4$, within 1 standard deviation of
the results for an isotropic DF.
Anisotropic models fitted to an NFW potential yield similar mass estimates but cannot constrain
the concentration radius $r_c$ in the range $10-500$ kpc although $r_c$ is as small
as $26_{-4}^{+27}$ kpc in isotropic NFW models.
Assuming a constant mass-to-light ratio, the derived $M/L$ in the B-band 
is $125\pm10\,\hbox{M}_{\sun}/\hbox{L}_{\sun}$, but
the model fit is substantially worse than for other mass models, and the derived mass-to-light ratio
is far larger than any plausible stellar population.
Our methods and results are more rigorous than
earlier attempts to
measure the mass of M87 from GC kinematics \citep{MT93, Coh97} and differ from those
in \citet{Rom01} as discussed in \S\ref{sec:dis}.  

We find that our mass estimates are insensitive to the smoothing parameters.
Smoothing has stronger effects on the DF, which is determined less
reliably. The
energy distribution has multiple peaks for the chosen
smoothing parameters, perhaps suggesting a few kinematically distinct groups of GCs in M87.
The derived anisotropy of the GCs can not be
determined at this stage. With more data, it may
be possible to measure the anisotropy of the velocity distribution and to test models of
GC evolution predicting that GCs on less-eccentric orbits are more likely to survive \citep[e.g.][]{Gne97}.

A more statistically robust method of choosing smoothing parameters,
more kinematic data
and a more accurate surface number density, especially beyond 110 kpc 
are the most important theoretical and observational advances needed to
determine the mass distribution in M87 more accurately by this method.
For example, the ongoing ACS Virgo cluster survey \citep{Cot04} will observe
20,000 globular clusters in the Virgo cluster, which may allow us
to put tighter constraints on the mass density profiles of M87 and other member galaxies.

\acknowledgements
We thank Judy Cohen for providing the data in Figures \ref{fig:fig4} and  \ref{fig:fig5} in tabular form,
Michael Strauss, David Spergel, James Gunn and Robert Lupton for valuable discussions,
and the referee for suggestions that significantly improved the paper.
This research was supported in part by NASA grant NNG04GL47G and used
computational facilities supported by NSF grant AST-0216105.

\appendix
\section{Calculation of ${\lowercase{g_\k(\uppercase{R}, v_z)}}$}
\label{sec:Appendix}
We divide $E-L$ space into $N_E\times N_L$
rectangular bins. 
Assume that the bin $\k$ is defined by
\begin{eqnarray}
E&\in&[E_{m-1}, E_{m}],\nonumber\\
L&\in&[L_{n-1}, L_{n}]. \label{eq:ELbin}
\end{eqnarray}
We wish to determine the normalized distribution $g_\k(R, v_z)$ in projected radius and radial
velocity corresponding to a uniform phase-space density in bin $\k$ (eq. \ref{eq:gkRvz}). For an anisotropic DF,
this requires a three-dimensional integration over $v_x$, $v_y$ and $z$, which is
computationally expensive.
We may achieve considerate speedup by integrating analytically over $v_y$ (recall from eq. \ref{eq:EL3D} that
we assume that each GC lies on the $x$--axis).
We have two constraints on $v_y$ from inequalities (\ref{eq:ELbin}). Based on equation (\ref{eq:EL3D}), we have
\begin{eqnarray}
v_y^2&\in   & [A, B] \nonumber \\
     &\equiv& \left [2(E_{m-1}-\Phi(r))-(v_x^2+v_z^2), 
     2(E_{m}-\Phi(r))-(v_x^2+v_z^2) \right ], \\
v_y^2&\in& [C, D] \nonumber\\
     &\equiv &\left [\frac{L_{n-1}^2-(v_zR-v_xz)^2}{(R^2+z^2)}, 
     \frac{L_{n}^2-(v_zR-v_xz)^2}{(R^2+z^2)} \right]. 
\end{eqnarray}

Then we may set
\begin{eqnarray}
I\equiv \max (\min (B, D),0) \;\;\;\; J\equiv \max (\max (A, C),0), 
\end{eqnarray}
so that we have
\begin{eqnarray}
\int dv_y=\max((\sqrt{J}-\sqrt{I}), 0).
\end{eqnarray}
This algorithm speeds up the integration by a factor of 20 or so.

\clearpage

\clearpage

\clearpage

\clearpage

\clearpage

\clearpage

\clearpage

\clearpage

\clearpage

\clearpage

\clearpage

\clearpage

\end{document}